\documentclass[traditabstract]{aa}

\usepackage{graphicx}
\usepackage{txfonts}
\usepackage{url}
\usepackage{natbib}
\usepackage{color}
\bibpunct{(}{)}{;}{a}{}{,}
\begin{document}
\title{Microwave and radio emission of dusty star-forming galaxies: Implication for the cosmic radio background}

\author{N. Ysard
\and G. Lagache}
\institute{IAS, CNRS (UMR8617), Universit\'e Paris-Sud 11, B\^atiment 121, F-91400 Orsay, France, \email{nathalie.ysard@ias.u-psud.fr}}

\abstract
{We use the most up-to-date cosmological evolution models of dusty star-forming galaxies and radio sources to compute the extragalactic number counts and the cosmic background from 408~MHz to 12~THz (or 24~$\mu$m). The evolution model of star-forming galaxies reproduces the recent constraints obtained by Spitzer, Herschel, and ground-based submm and mm experiments: number counts, redshift distribution of galaxies, cosmic background intensity and anisotropies. The template spectral energy distributions used in this model are extrapolated to the radio domain adding three emission components: synchrotron, free-free, and spinning dust. To fix the synchrotron intensity, we use the well-known IR/radio flux ratio, $q_{70}$, and a constant spectral index $\beta_S = -3$, consistent with measurements made in local galaxies taking account the spinning dust emission.
For a constant $q_{70}$, our model added to the AGN contribution provides a good fit to the extragalactic number counts from 24 $\mu$m to 408 MHz, and to the cosmic background intensity in the far- and mid-IR. The spinning dust emission accounts for up to 20\% of the cosmic microwave background produced by star-forming galaxies, but for only less than 10\% of the total background when AGN are included. The star-forming galaxies account for 77.5\% of the number counts at 1.4~GHz for a flux of 100~$\mu$Jy. However, the model falls short of reproducing the cosmic radio background measured with the ARCADE2 balloon-borne experiment. Considering the case when $q_{70}$ decreases strongly with redshift, this still does not explain the ARCADE2 measurements. It also yields to an overestimate of the low-flux number counts in the radio. As a result, we rule out a steep variation of $q_{70}$ with the redshift at least for $z \leqslant 3.5$. Then, adding a population of faint star-forming galaxies at high redshift ($L_{IR} \leqslant 10^{11} L_{\odot}$ and $4 \leqslant z \leqslant6$), which would be able to reproduce the ARCADE2 measurements, leads to predictions of the cosmic IR background much higher than what is currently observed, ruling out this as the explanation for the ARCADE2 results.
Considering our findings and previous studies of the diffuse extragalactic radio emission, we conclude that if the radio emission measured by ARCADE2 is astrophysical in origin, it has to originate in the Galaxy or to originate in a new kind of radio sources (with no mid- to far-IR counterparts) or emission mechanism still to be discovered.}

   \authorrunning{}
\titlerunning{Microwave and radio emission of star-forming galaxies}

\maketitle

\section{Introduction}
\label{section_introduction}

The extragalactic background light or cosmic background is dominated by the Cosmic Microwave Background (CMB), a relic of the evolution of the early Universe. The remainder of the background is directly linked to the formation and evolution of all the galaxies in the Universe apart from our Galaxy. Even if less energetic, this background can teach us a lot about galaxy evolution throughout time and can be separated into five components: the X- and $\gamma$-ray backgrounds, the cosmic optical background, the Cosmic Infrared Background (CIB),and the Cosmic Radio Background (CRB). The CIB is produced by the emission of dust illuminated by the stars contained in infrared (IR) or dusty star-forming galaxies. The evolution of these galaxies is well constrained by direct measurements of the CIB, galaxy number counts from the mid- to the far-IR, and redshift distribution of galaxies. Star-forming galaxies are also expected to emit at radio frequencies because the stars they contain can ionise their gas content and thus produce free-free emission. Eventually, these stars will evolve into supernovae leading to synchrotron emission. Anomalous microwave emission (AME), assumed to be produced by spinning dust grains \citep{PlanckDickinson2011, PlanckMarshall2011}, may also arise from dusty star-forming galaxies. The exact intensity and evolution of radio emission from star-forming galaxies is still a matter of debate.

Recent measurements of the CRB made by \citet{Fixsen2011} with the ARCADE2 experiment have renewed the interest in the contribution of star-forming galaxies at microwave wavelengths. Indeed, \citet{Fixsen2011} measured a background between 3 and 10~GHz that is at least $5\sigma$ above the Cosmic Microwave Background (CMB) temperature estimated by COBE/FIRAS. Investigations made by \citet{Singal2010} showed that standard extragalactic radio sources cannot account for this radio excess and they concluded that it has to come from star-forming galaxies. However, up till now, no extragalactic source evolution model has been able to explain this excess \citep{Massardi2010, Singal2010, Vernstrom2011, Ponente2011, Draper2011}. Our aim is to provide a comprehensive evolution model of star-forming galaxies able to reproduce all the observations, cosmic background intensity and extragalactic number counts, from the mid-IR to the radio. To do so, we use the most up-to-date galaxy evolution models for both star-forming galaxies \citep{Bethermin2011} and radio-loud AGN \citep{deZotti2005, Massardi2010, Tucci2011}. Then, we test several hypothesis to try to explain the excess measured by the ARCADE2 team: extragalactic spinning dust emission? Variation of the well-known far-IR/radio flux ratio with redshift? Additional population of faint star-forming galaxies at high redshift?

The paper is organised as follows. Section \ref{section_model} describes the evolution model of dusty star-forming galaxies we use, as well as the emission components added to explain radio emission (synchrotron, free-free, spinning dust). The models for the radio-loud AGN are described in Sect. \ref{section_AGN}. Our results are then presented in Sect. \ref{section_results}. Finally we give our conclusions in Sect. \ref{section_conclusions}.

\section{Evolution model for dusty star-forming galaxies}
\label{section_model}

\subsection{Luminosity function evolution and IR spectral energy distributions}
\label{luminosity_evolution}

We used the recent model of \citet{Bethermin2011} to compute the number counts of dusty star-forming galaxies at the radio wavelengths. Two ingredients come into play: the spectral energy distribution (SED) of galaxies and the luminosity function (LF) evolution.

The SEDs, from 4~$\mu$m to 1~mm, are from the template library of \cite{Lagache2004}, which consists of two galaxy populations. First, a star-forming galaxy population with SEDs that vary with IR bolometric luminosities. Second, a normal-galaxy population with a fixed template SED that is colder than the star-forming galaxy templates. The normal and star-forming galaxies are dominant at low- and high-luminosity, respectively. The fraction of each galaxy population as a function of the bolometric luminosity is given by a smooth function
\begin{equation}
\label{eq:mixpop}
\frac{\Phi_{\rm starburst}}{\Phi} = \frac{1+th \bigl [ {\rm log}_{10}(L_{\rm IR}/L_{\rm pop}) /\sigma_{\rm pop} \bigl ]}{2},
\end{equation}
where $th$ is the hyperbolic tangent function, $L_{\rm pop}$ the luminosity at which the number of normal and star-forming galaxies are equal, and $\sigma_{\rm pop}$ characterises the width of the transition between the two populations. At $L_{\rm IR}=L_{\rm pop}$, the starbursts fraction is 50\%.

\citet{Bethermin2011} assume that the luminosity function (LF) is a classical double exponential function:
\begin{equation}
\label{eq:lf}
\Phi(L_{\rm IR}) = \Phi^\star \, \times \left( \frac{L_{\rm IR}}{L^\star} \right)^{1-\alpha} \,  \times \,exp \left[ -\frac{1}{2\sigma^2} {\rm log}_{10}^2 \left(1+\frac{L_{\rm IR}}{L^\star}\right) \right],
\end{equation}
where $\Phi(L_{\rm IR})$ is the number of sources per logarithm of luminosity, and per comoving volume unit for an infrared bolometric luminosity $L_{\rm IR}$. $\Phi_{\star}$ is the normalisation constant characterising the density of sources, $L_{\star}$ is the characteristic luminosity at the break, and $1-\alpha$ and $1-\alpha-1/\sigma^2/ln^2(10)$ are the slopes of the asymptotic power-law behaviour at low- and high-luminosity respectively.
To reproduce the observations in the infrared, \citet{Bethermin2011} allow the parameters of the LF to vary with redshift. In order to limit the number of free parameters and to avoid strong degeneracies, $\alpha$ and $\sigma$ are fitted as constants with $z$. A continuous LF redshift-evolution in luminosity and density is assumed following broken power-laws: $L^\star \propto (1+z)^{r_L}$  and $\Phi^\star \propto (1+z)^{r_\Phi}$, where $r_L$ and $r_{\phi}$ are parameters driving the evolution in luminosity and density, respectively. It is impossible to reproduce the evolution of the LF with constant $r_L$ and $r_\phi$. \citet{Bethermin2011} consequently allow their value to change at two specific redshifts: three different values of $r_L$ and $r_{\phi}$ are used at low redshift ($0 < z < z_{break}$), intermediate redshift ($z_{break} < z < 2$), and high redshift ($z > 2$). The position of the first redshift break is a free parameter and converges to the same final value ($z_{break} \sim 0.9$) for initial values $0 < z < 2$. The second break does not appear naturally in the fitting procedure and is fixed at $z = 2$\footnote{Numerical values and details about the fitting procedure are given in \citet{Bethermin2011}.}.

The model has 13 free parameters that were determined by fitting the model to published measurements of galaxy number counts and monochromatic LF measured at given redshifts:
\begin{itemize}
\item Number counts: {\it Spitzer} counts at 24, 70 and 160~$\mu$m, {\it Herschel} counts at 250, 350 and 500~$\mu$m, and AzTEC counts at 1.1~mm.
\item Monochromatic LFs: IRAS local LF at 60~$\mu$m, {\it Spitzer} LF at 24~$\mu$m at z=0, at 15~$\mu$m at z=0.6, at 12~$\mu$m at z=1, and at 8~$\mu$m at z=2.
\item FIRAS CIB spectrum between 200~$\mu$m and 2~mm.
\end{itemize}
The best-fit parameters, as well as their uncertainties and degeneracies, were obtained using a Monte Carlo Markov chain (MCMC) Metropolis-Hastings algorithm. The model is adjusted to reproduce deep counts and monochromatic LFs at key wavelengths. It also reproduces recent observations that constrain very well the models, such as the \cite{Jauzac2011} measured redshift distribution of the CIB.

We used the so-called {\it mean model}, which is obtained using the mean value of the parameters as given in \citet{Bethermin2011} with the lensing contribution that is available on the \url{http://www.ias.u-psud.fr/irgalaxies/} web page.

The SED templates of \citet{Lagache2004} used in \citet{Bethermin2011} stop at $\lambda = 1$ mm. To predict the number counts and cosmic background at radio wavelengths of dusty star-forming galaxies, we need to extrapolate the SED templates into the radio domain. We consider three emission components: synchrotron (Sect. \ref{section_synchrotron}), free-free (Sect. \ref{section_free_free}), and spinning dust (Sect. \ref{section_spinning}).

\subsection{Synchrotron}
\label{section_synchrotron}

Synchrotron emission is produced by relativistic electrons interacting with a magnetic field. At first order, its emissivity can be expressed as a power-law: $\epsilon_{S}(\nu) \propto \nu^{\beta_S}$. Using WMAP intensity and polarisation data, \citet{MAMD2008} measured a spectral index $\beta_S = -3.0 \pm 0.06$ in our Galaxy between 408~MHz and 23~GHz, in agreement with previous measurements by \citet{Giardino2002} and \citet{Platania2003}. In the nearby spiral galaxy M33, \citet{Tabatabaei2007} showed that, even if the synchrotron spectrum hardens from the outer parts of the galaxy to the star-forming regions, a constant emissivity spectral index, $\beta_S = -2.953 \pm 0.285$, is appropriate for global studies. Similar values were found by \citet{Peel2011} in three nearby star-forming galaxies  (M82, NGC253, and NGC4945). Consequently, and for the sake of simplicity, we adopt a single power-law for the two galaxy templates of \citet{Lagache2003} with $\beta_S = -3.0$.

The radio continuum and far-IR emissions are observed to be tightly correlated in external galaxies \citep{Helou1985}. This correlation is remarkable in the sense that it holds over five orders of magnitude of radio continuum luminosity \citep{Yun2001}, and both at the scale of an entire galaxy and at subgalactic scales \citep{Tabatabaei2007, Hughes2006}. The correlation is also observed in the Milky Way \citep{Boulanger1988} at large scale and down to a few parsecs scale \citep{Zhang2010}. \citet{Boulanger1988} and \citet{Zhang2010} found that the far-IR/radio flux ratio, usually called $q$, has similar values in the Milky Way and in external galaxies. Because synchrotron is the dominant emission component at 1.4~GHz, we set its intensity using the far-IR/radio flux ratio, $q_{70} = \log(S_{70 \, \mu{\rm m}} / S_{1.4 \, {\rm GHz}})$, of IR emission at 70~$\mu$m and radio emission at 1.4~GHz.

The possible evolution of $q_{70}$ with redshift is still debated, some authors finding a constant relation \citep{Ibar2008, Sargent2010}, where others find a slow decrease with redshift \citep{Seymour2009, Beswick2008, Ivison2010}. As a result, we will consider two cases in this paper. The first one assumes that the far-IR/radio flux ratio is constant over redshift and equal to $q_{70} \sim 2.15$ \citep{Appleton2004, Frayer2006, Seymour2009}. In the following, we will refer to it as case $A$. The resulting spectra for the normal spiral galaxy and starbust galaxy templates are displayed in Fig.~\ref{figure_templates}. The second one assumes that the ratio evolves with redshift as $q_{70}(z) \sim 2.15 - 0.72 \log(1+z)$. This relation comes from \citet{Seymour2009} and is the steepest that has been reported so far. We choose it to maximize the effect of varying $q_{70}$ on extragalactic number counts. We will refer to it as case $B$. Case $B$ will lead to a brighter synchrotron emission for high redshift and should consequently have a stronger impact on the prediction of the CRB than case $A$.

\subsection{Free-free}
\label{section_free_free}

The free-free component is particularly difficult to estimate as it is never dominant at any wavelength. According to \citet{Peel2011} and \citet{Murphy2010}, we model the free-free spectrum as a power-law with a spectral index $\beta_{FF} = -2.1$. This value is similar to what is measured in our Galaxy \citep{MAMD2008, Dickinson2003}. In previous studies, before anomalous microwave emission (or spinning dust) was observed in galaxies, the free-free intensity was often normalised to be equal to that of the synchrotron at about 30~GHz \citep{Condon1992, Ponente2011}. This means that spinning dust emission was included in the free-free and/or synchrotron emission. However, the spectral shape of the spinning dust emission is a "bump" centred around 20-30~GHz while the free-free and synchrotron emissions span over a much broader range of frequencies. Thus, ignoring the spinning dust contribution can lead to a flatter spectral index for the free-free and/or synchrotron emissions that implies an overestimate of their intensity. Consequently, we use recent results taking into account spinning dust emission to normalise the free-free emission at 10~GHz. For the three nearby star-forming galaxies observed by \citet{Peel2011}, the free-free emission is about half of that of the synchrotron at 10 GHz. Similar values are found in NGC6946 by \citet{Murphy2010}: the free-free intensity reaches about (52 $\pm$ 12)\% of the synchroton intensity at 10 GHz throughout the galaxy. These values also match the observations of \citet{Tabatabaei2007} for M33. Consequently, we normalise the intensity of the free-free emission to be equal to 50\% of the synchrotron at 10 GHz. The resulting spectra for the normal spiral galaxy and star-forming galaxy templates are displayed in Fig. \ref{figure_templates} for case $A$.

\subsection{Spinning dust emission}
\label{section_spinning}

Anomalous microwave emission, peaking between 10 and 60 GHz, is observed everywhere in our Galaxy, from the Galactic plane \citep{PlanckMarshall2011} to intermediate Galactic latitudes \citep{MAMD2008, Davies2006, LagacheAME2003}. AME is detected in various interstellar clouds from the most diffuse to dense molecular clouds \citep{PlanckDickinson2011, Vidal2011, Casassus2008, Casassus2006, Watson2005}. Up to now, a few extragalactic detections have been reported in nearby star-forming galaxies \citep{Peel2011, Murphy2010, Scaife2010} and in the Small Magellanic Cloud \citep{PlanckBernard2011, Bot2010}. It is usually assumed that AME is due to the emission of rapidly rotating interstellar PAHs, the so-called spinning dust emission, as first suggested by \citet{DL98}. The rotation of the PAHs is driven by interactions with the interstellar gas particles and absorptions of UV/visible stellar photons. This means that any environment containing PAHs should emit in the microwave. The spectra of the dusty star-forming galaxies responsible for the CIB exhibit bands in the mid-IR ($\sim$ 3 to 20 $\mu$m) characteristic of the interstellar PAHs. Consequently, we add a spinning dust component to our galaxy templates to estimate its impact on the source counts and cosmic background intensity in the microwave and radio wavelengths.

The intensity and the peak frequency of the spinning dust spectrum depend on the strength of the interstellar radiation field heating the grains and on the local gas density and temperature \citep{DL98, Ysard2011}. Thus, two parameters have to be set to determine the spinning dust component for star-forming galaxies: a mixture of environments and the radiation field intensity.

For the Milky Way, the component separation performed by \citet{MAMD2008} using WMAP data allowed \citet{Ysard2010} to measure a ratio between the AME in the WMAP 23 GHz band and the PAH emission in the IRAS 12 $\mu$m band of $L_{23 \, {\rm GHz}}/L_{12 \, \mu{\rm m}} = (5.36 \times 10^{-3}) \pm (4 \times 10^{-5})$ for the entire Galaxy. This ratio is valid for the radiation field illuminating our Galaxy. The intensity of the radiation field can be expressed in units of $G_0$-factor, equal to 1 in the case of the interstellar standard radiation field (ISRF) illuminating the solar neighbourghood for which \citet{Boulanger1996} and \citet{Lagache1998} measured an average dust temperature of $\sim 17.5$~K. Using the FIRAS spectrum and the DIRBE 100 $\mu$m data averaged over the entire Galaxy, we determine the mean dust temperature of the Milky Way by fitting a modified blackbody with a constant spectral index $\beta=2$ for $100 \leqslant \lambda \leqslant 250 \; \mu$m. We find a dust temperature equal to $T_{SED} \sim 18$ K leading to $G_0^{{\rm MW}} \sim (T_{SED}/17.5 \; {\rm K})^{4+\beta} = 1.2$ \citep{Ysard2010}. Then, we proceed to the same fitting procedure to get estimates of the intensities of the mean radiation fields required to explain the far-IR spectra of our galaxy templates. For normal spiral galaxies, we get $G_0 \sim 2$ ($T_{SED} = 19.5$ K), and for star-forming galaxies with bolometric luminosities of $L_{IR} = 10^{10}, \, 10^{11}$, and $10^{12} \; L_{\odot}$, we get $G_0 \sim 13$, 22, and 45, respectively ($T_{SED} \sim 27$, 29, and 33 K, respectively). Note that the exact $\beta$ value and wavelength range used to compute the temperature are not critical as long as the same parameters are used for both the Galactic spectrum (the reference) and the extragalactic SED templates.

Then, the spectral shape of the spinning dust spectrum is determined with the spinning dust model described in \citet{Silsbee2011}. We assume that the ISM of galaxies can be approximated with a mixture of environments \citep{Dopita2003, Cox2005} composed by: 20\% of Warm Ionised Medium (WIM, with a proton density of $n_H = 0.1$ cm$^{-3}$), 20\% of Warm Neutral Medium (WNM, $n_H = 0.4$ cm$^{-3}$), 30\% of Cold Neutral Medium (CNM, $n_H = 30$ cm$^{-3}$), and 30\% of Molecular Cloud (MC, $n_H = 300$ cm$^{-3}$), and illuminated by the ISRF scaled with the $G_0$ values determined previously. Then, we integrate the galaxy template spectra in the IRAS 12 $\mu$m filter in order to normalise them to the $L_{23 \, {\rm GHz}}/L_{12 \, \mu{\rm m}}$ ratio measured by \citet{Ysard2010} in the Milky Way. To take into account the lower $G_0^{{\rm MW}} = 1.2$ value measured in the Milky Way, we multiply the spinning dust spectrum by a factor equal to the ratio of the integrated intensity in the WMAP 23~GHz band of this spinning dust spectrum computed for a given $G_0$ value and the same but computed for $G_0^{{\rm MW}} = 1.2$. This factor is equal to 1.00 in the case of normal spiral galaxies, and to 1.15, 1.28, and 1.45 for star-forming galaxies with bolometric luminosities of $L_{IR} = 10^{10}, \, 10^{11}$, and $10^{12} \; L_{\odot}$, respectively.

\begin{figure}[!t]
\centerline{
\includegraphics[width=0.4\textwidth]{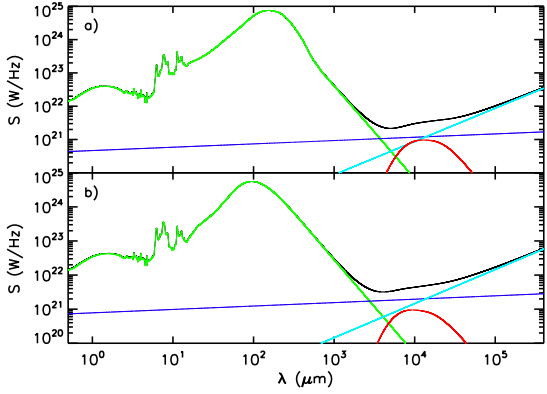}}
\caption{SED template for a galaxy with a bolometric luminosity of $10^{11} \, L_{\odot}$ (black line), contribution of IR dust emission (green line), free-free emission (blue line), synchrotron emission calculated for case $A$ (cyan line), and spinning dust emission (red line). a) Normal spiral galaxy template. b) Star-forming galaxy template.}
\label{figure_templates} 
\end{figure}

The resulting spectra are displayed in Fig. \ref{figure_templates} for case $A$ (constant $q_{70}$). For normal spiral galaxies, the spinning dust emission at 23 GHz accounts for 31.7\% of the total intensity. For star-forming galaxies with $L_{IR} = 10^{10}, \, 10^{11}$, and $10^{12}$~$L_{\odot}$, it accounts for 25.3\%, 18.5\%, and 12.8\% of the total intensity, respectively. If $q_{70}$ varies with the redshift (case $B$), the spinning dust component is then fully dominated by the synchrotron and free-free emissions.

\section{Evolution model for the AGN-powered radio sources}
\label{section_AGN}

The extragalactic radio number counts are dominated by radio-loud active galactic nuclei (AGN) for bright flux densities, $S \gtrsim 1$~mJy \citep{Ibar2008, Condon1992}. These radio-loud AGN are also expected to account for about half of the surface brightness of the CRB \citep{Massardi2010}. To take into account the contribution of radio galaxies, we use two different models, depending on the frequency. For $\nu = 408$~MHz, 610~MHz, 1.4~GHz, 2.7~GHz, 5~GHz, 15.7~GHz, 30~GHz, 44~GHz, and 70~GHz, we use the model of \citet{Massardi2010}. For $\nu = 8.44$~GHz, 20~GHz, 100~GHz, 143~GHz, and 214~GHz, we use the model C2Ex of \citet{Tucci2011}.

These two models provide good fits to the extragalactic number counts in the radio but the model of \citet{Tucci2011} gives a better repartition of the counts between the flat- and steep-spectrum galaxies. The model of \citet{Massardi2010} assumes two flat-spectrum populations with different luminosity functions and one steep-spectrum population. For each population, they adopt a power-law spectrum $S_{\nu} \propto \nu^{-\alpha}$ with $\alpha = 0.1$ or $\alpha = 0.8$. The model of \citet{Tucci2011} allows for a gaussian distribution for the spectral indices. They also introduce an additional population with an inverted spectrum and define frequency breaks for the slopes of the flat-spectrum populations.

\section{Results and discussion}
\label{section_results}

\subsection{Constant $q_{70}$ and spinning dust emission (case $A$)}

\begin{figure*}[!t]
\centerline{
\begin{tabular}{ccc}
\includegraphics[width=0.32\textwidth]{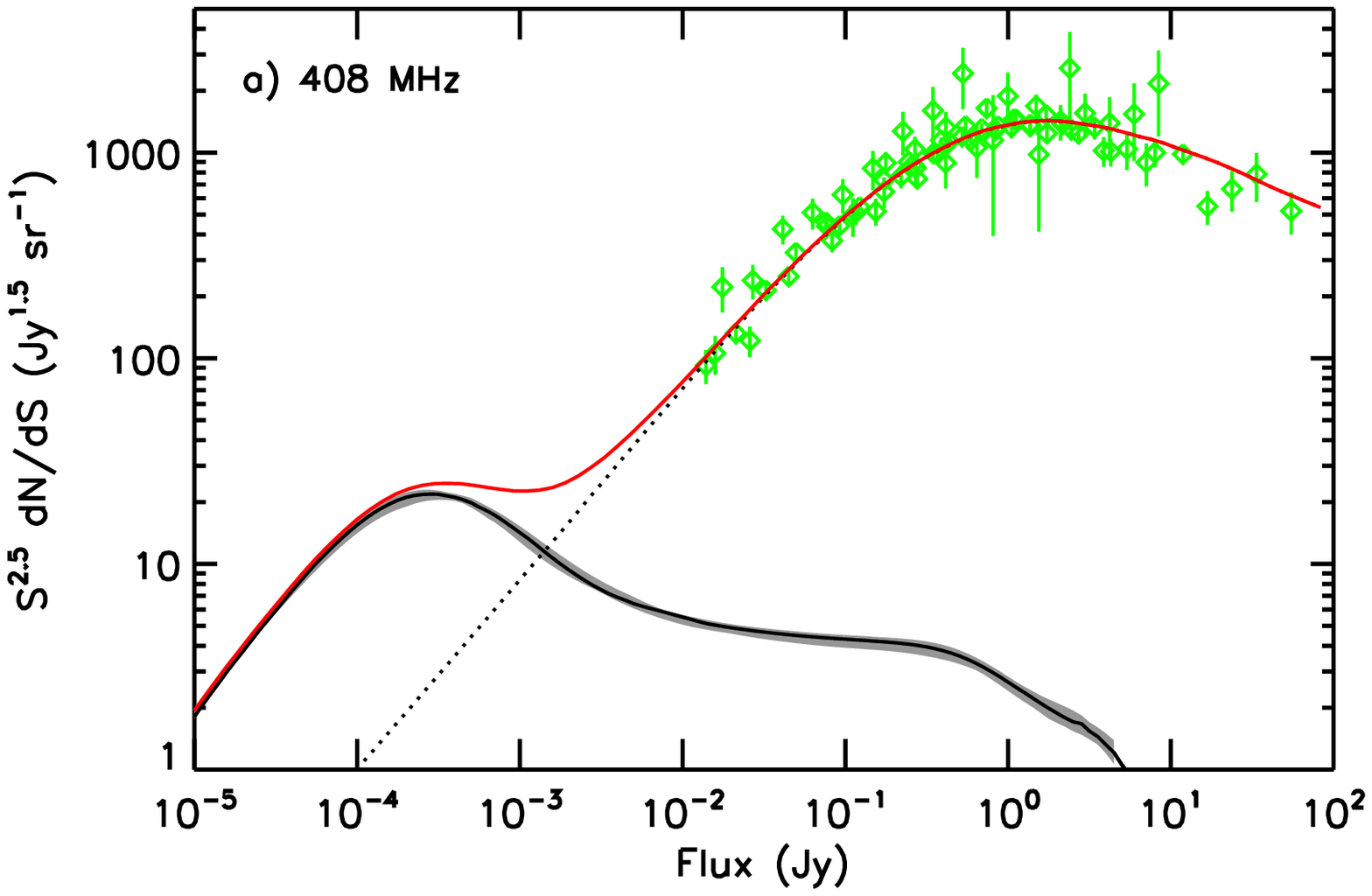} & \includegraphics[width=0.32\textwidth]{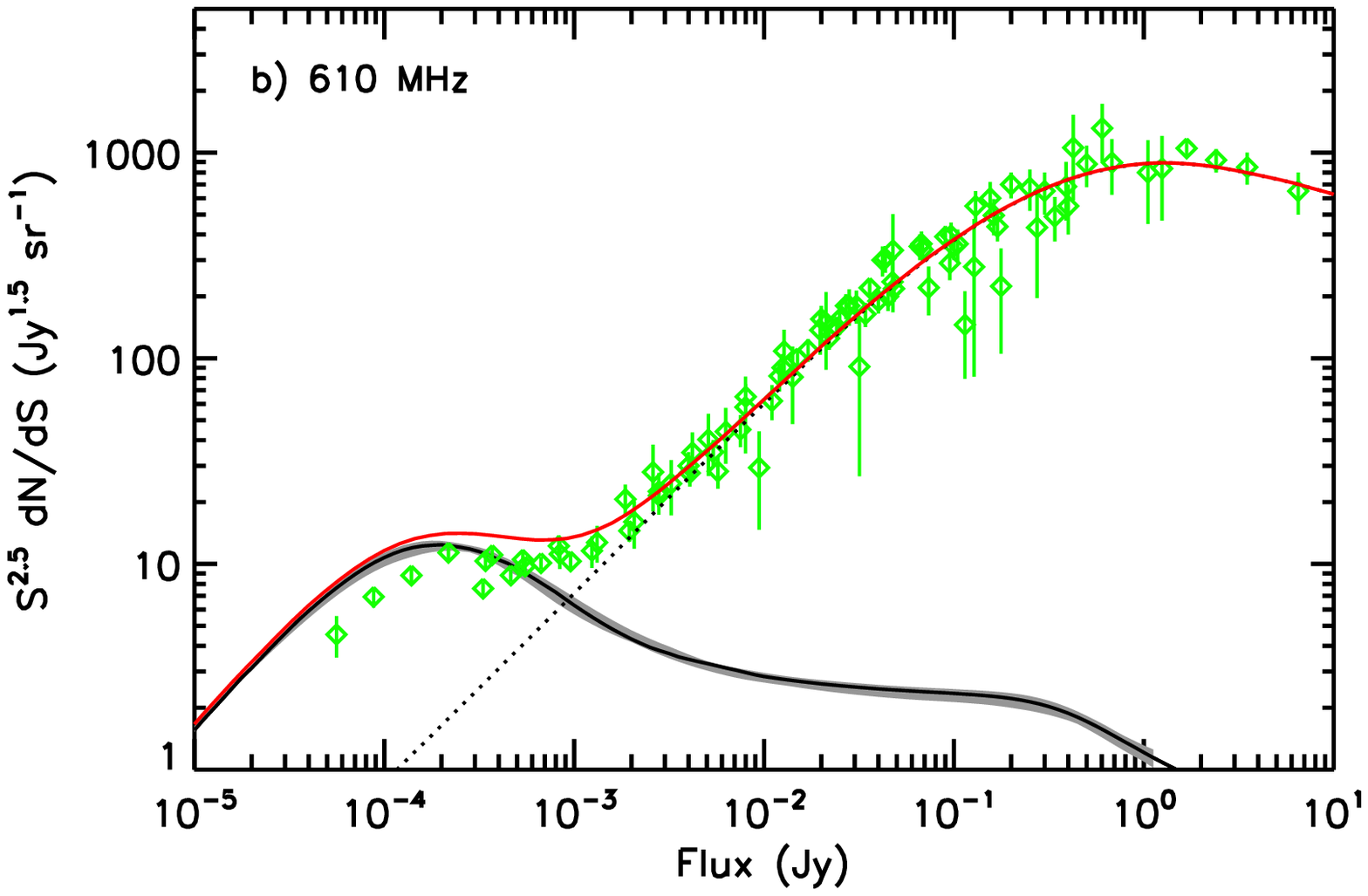}  & \includegraphics[width=0.32\textwidth]{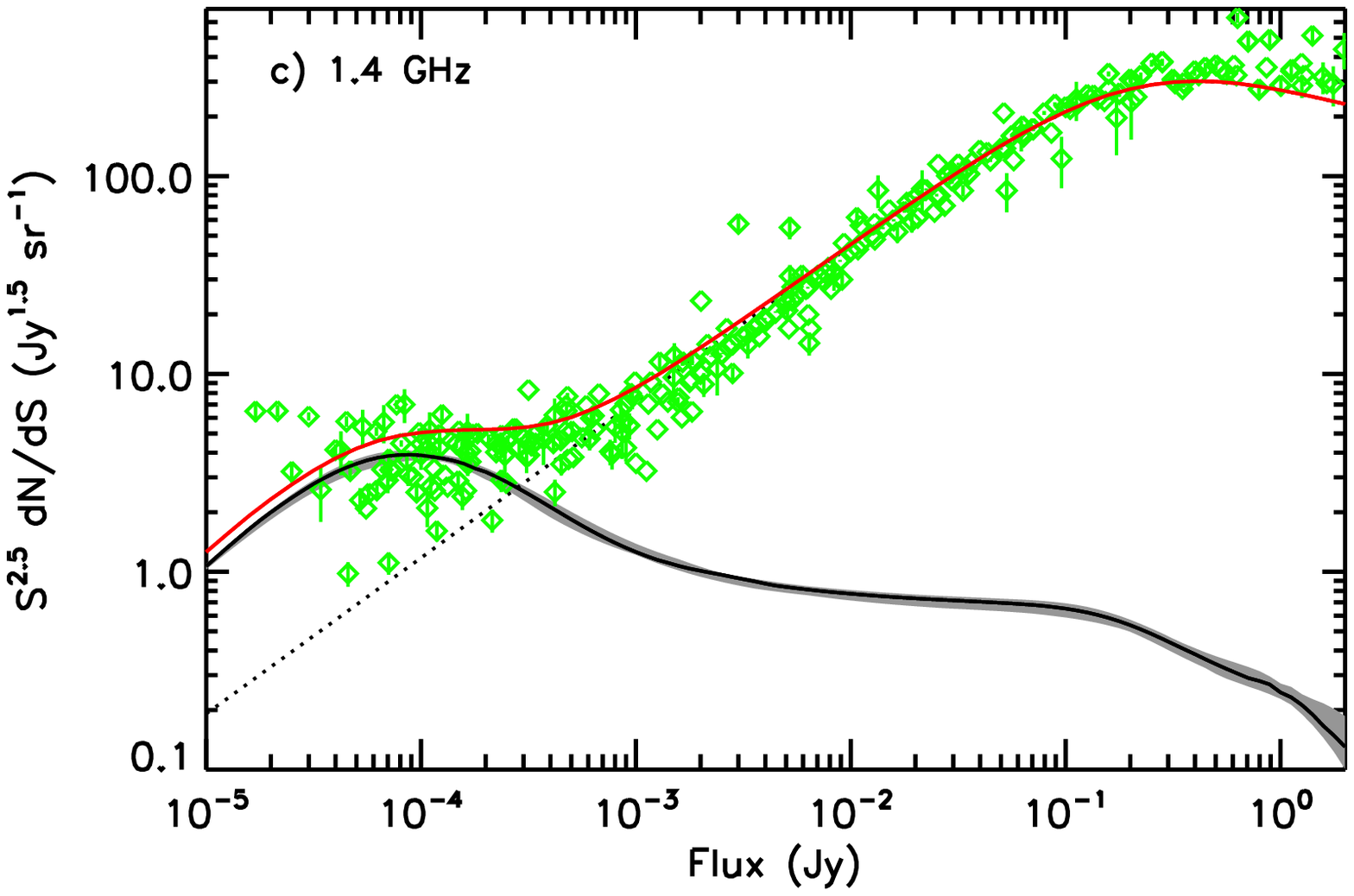} \\
\includegraphics[width=0.32\textwidth]{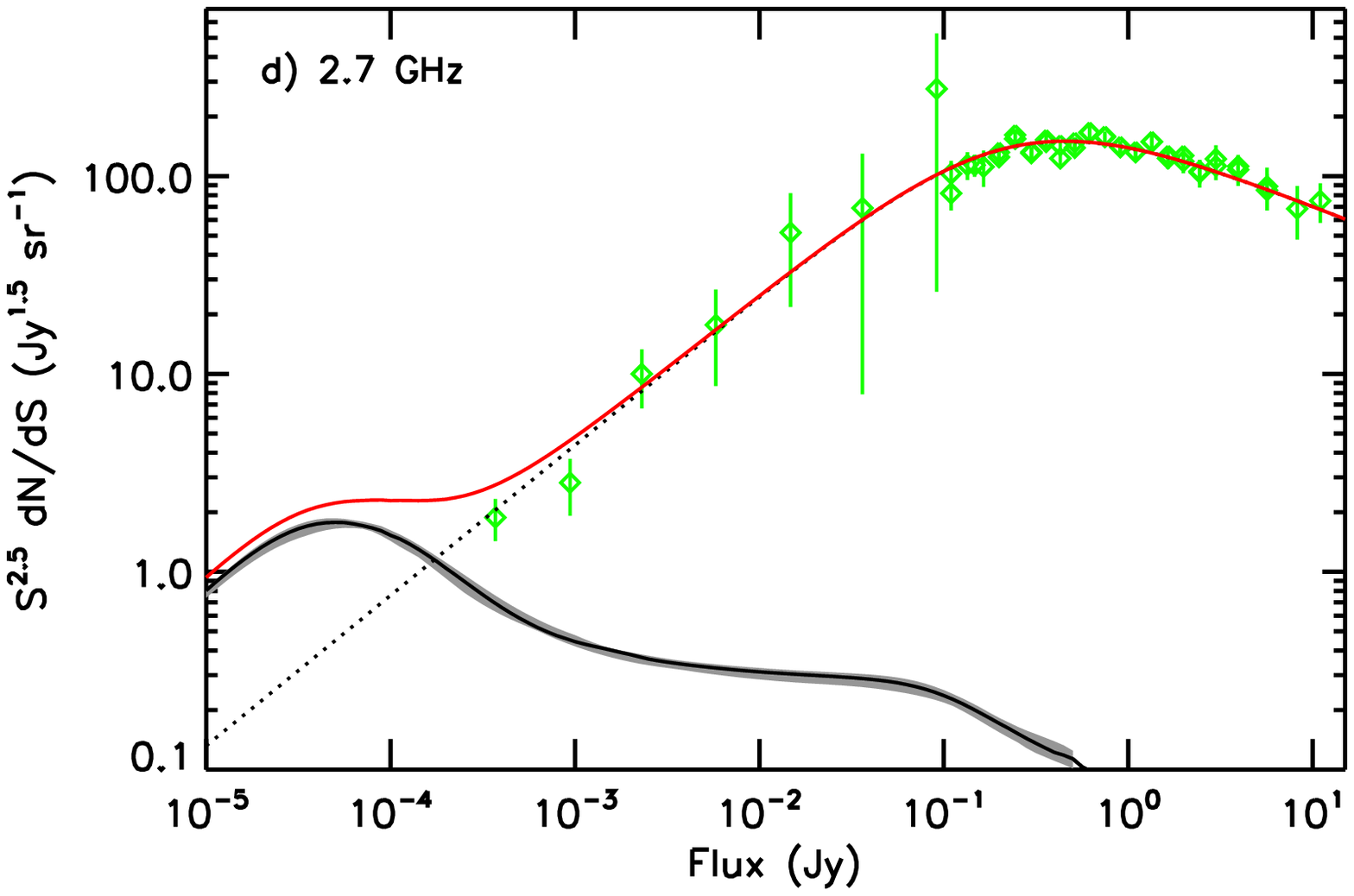} & \includegraphics[width=0.32\textwidth]{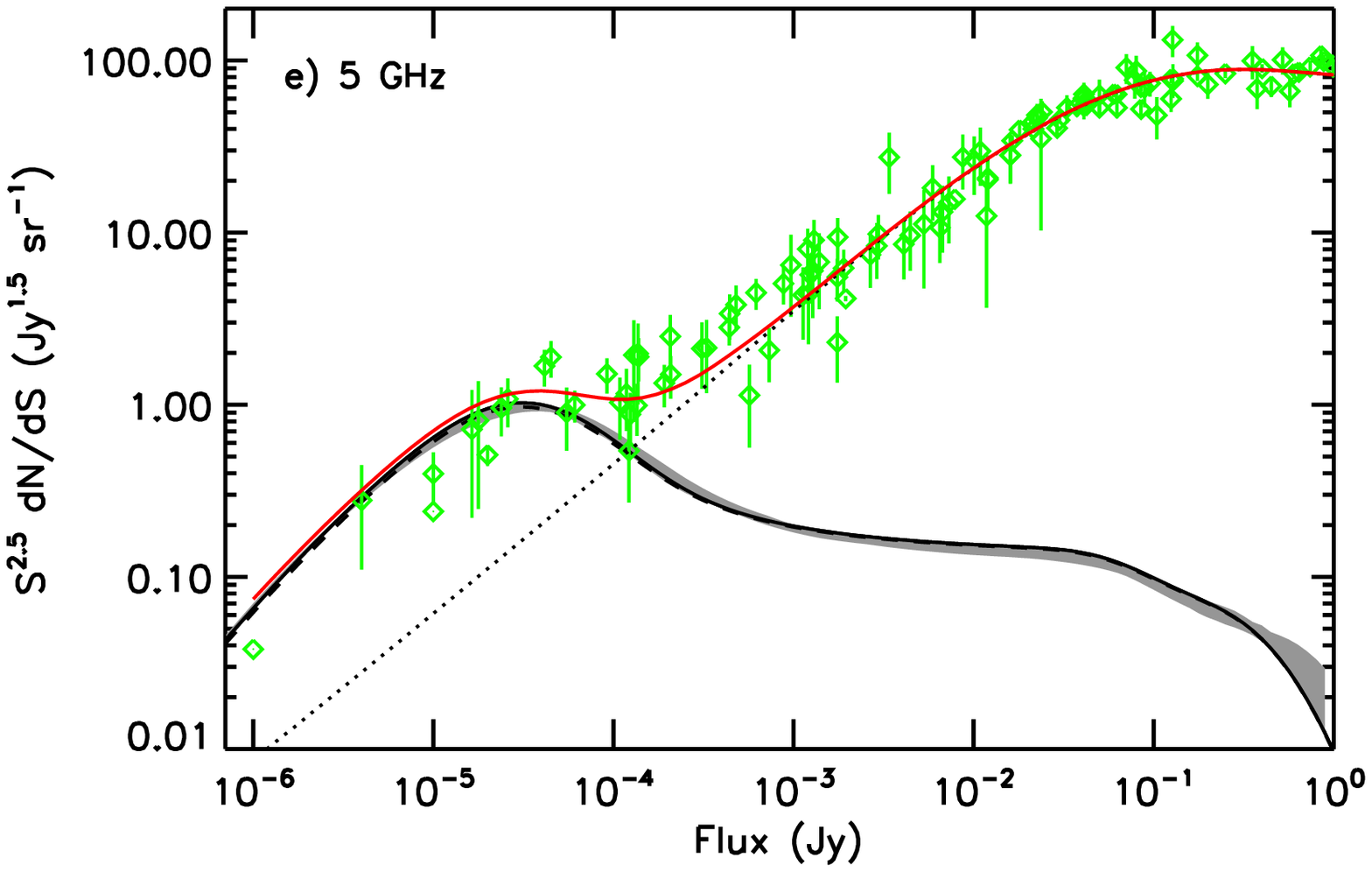} & \includegraphics[width=0.32\textwidth]{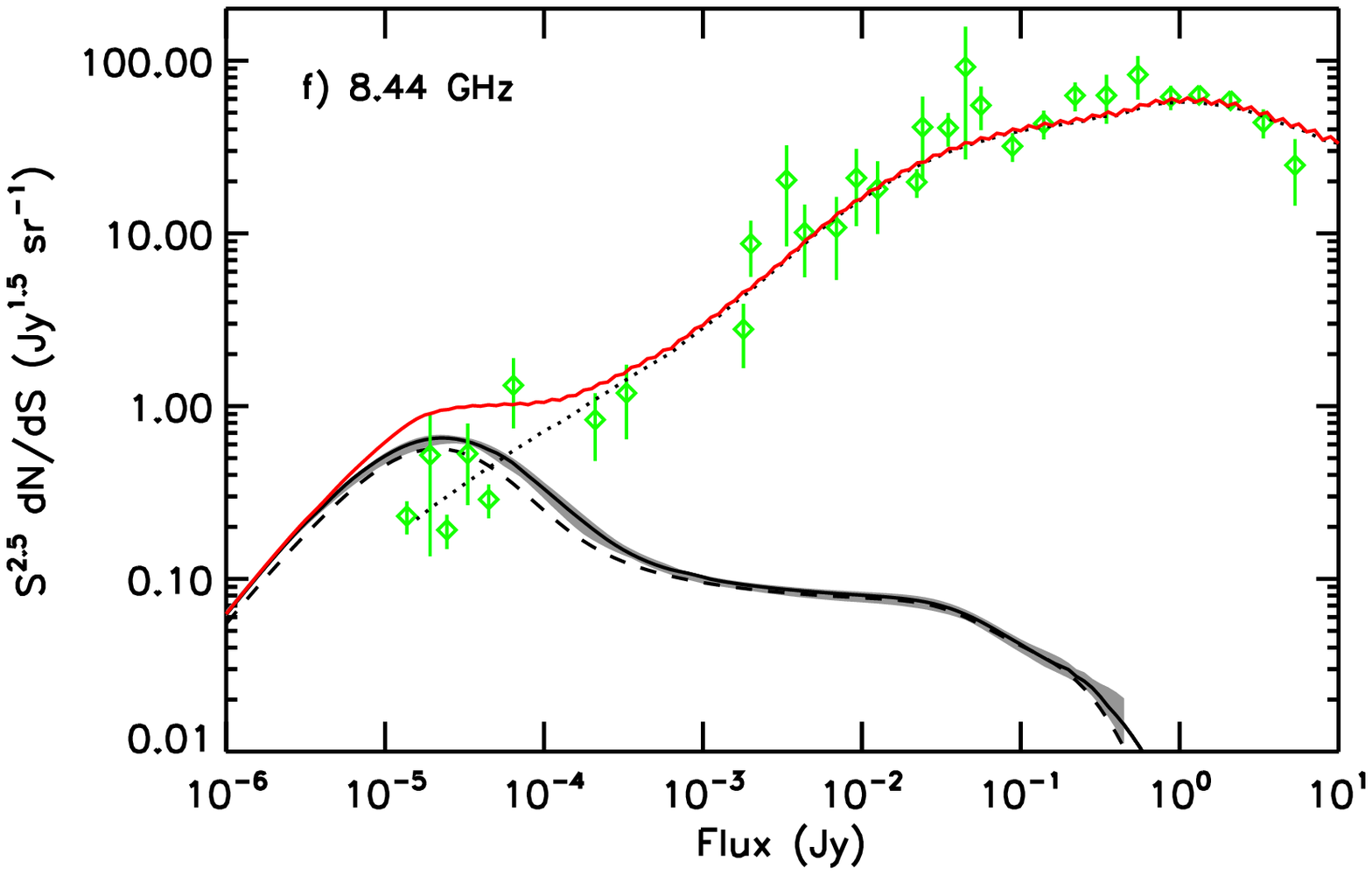} \\
\includegraphics[width=0.32\textwidth]{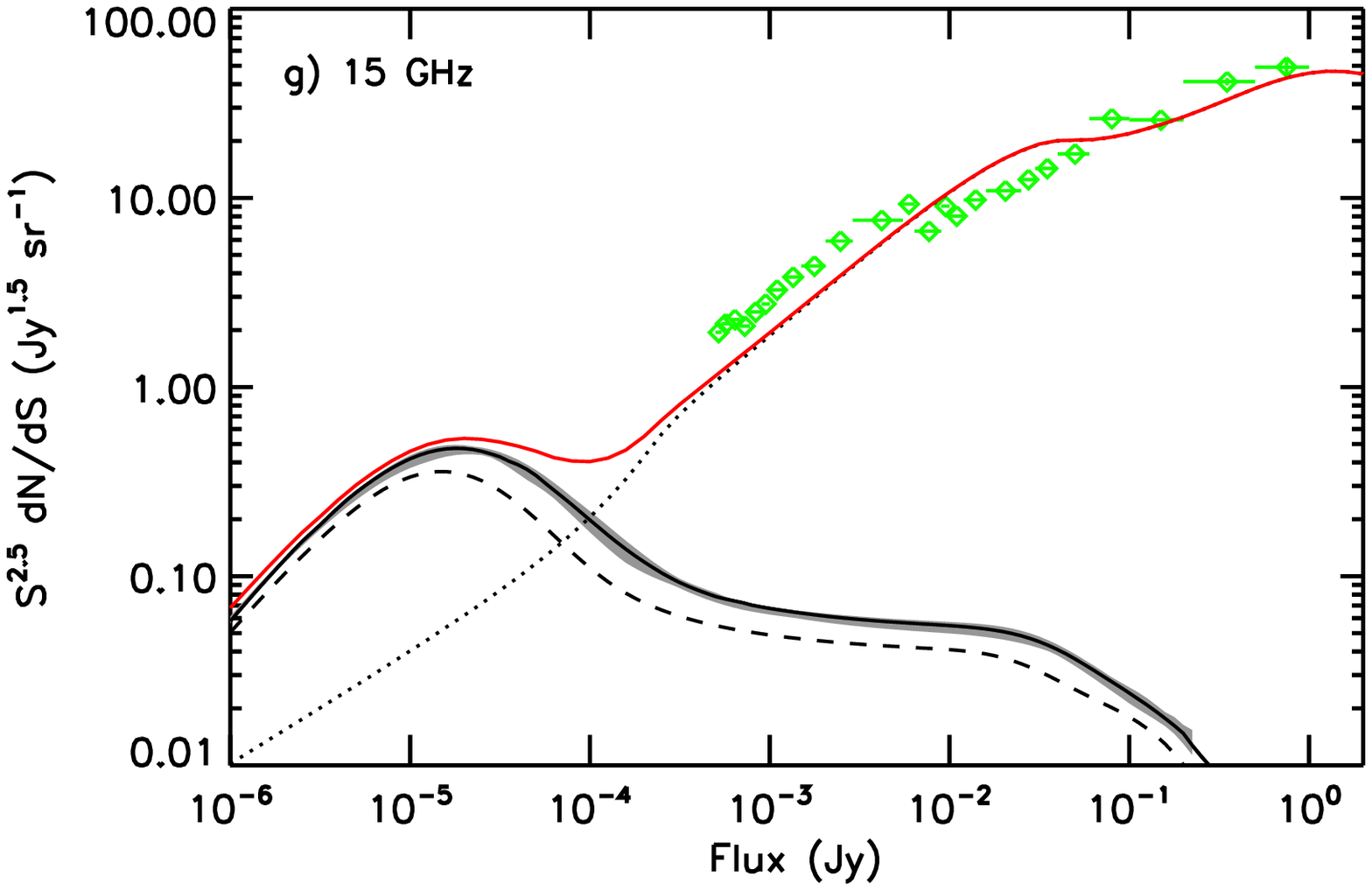} & \includegraphics[width=0.32\textwidth]{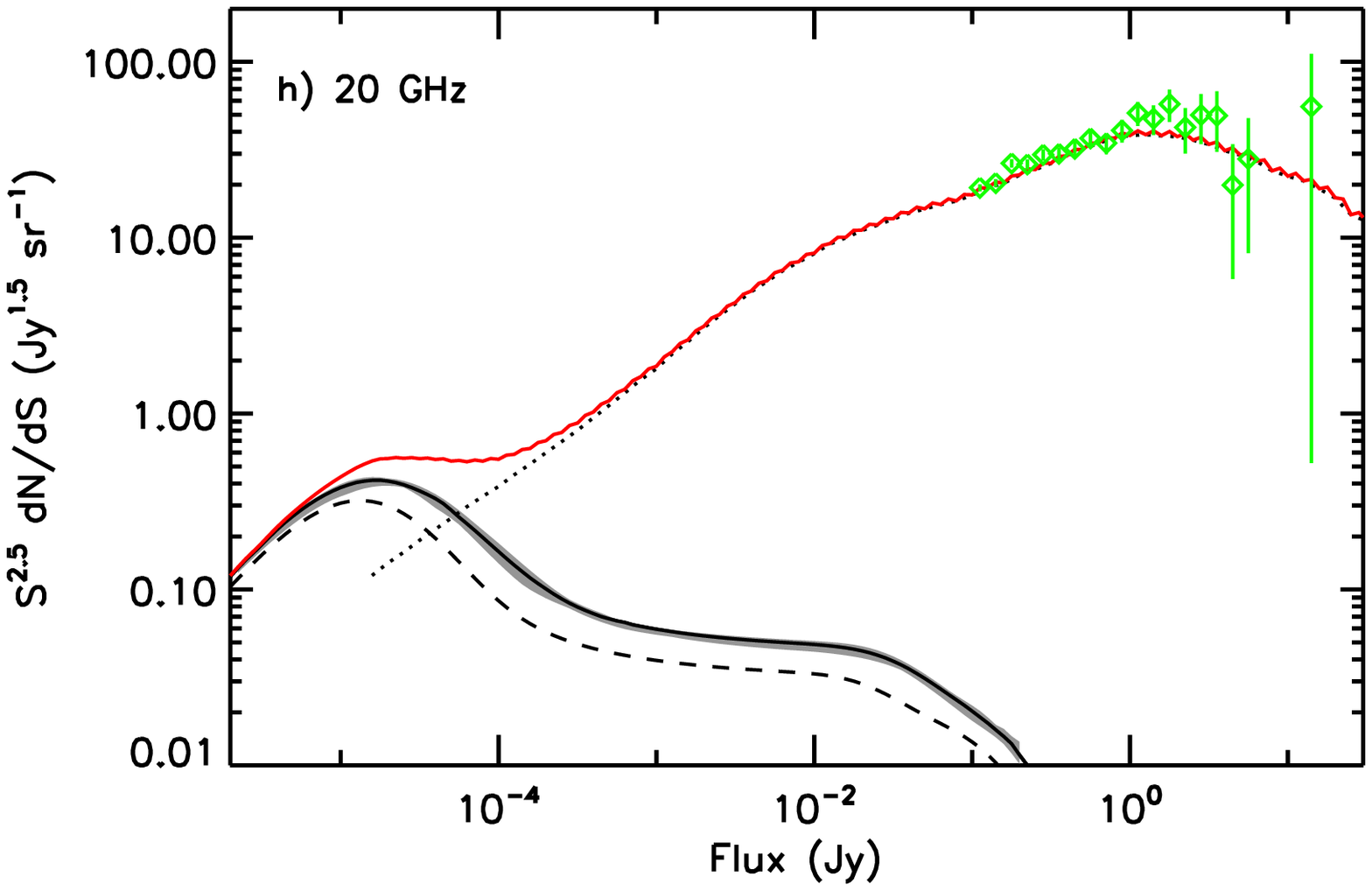} 
& \includegraphics[width=0.32\textwidth]{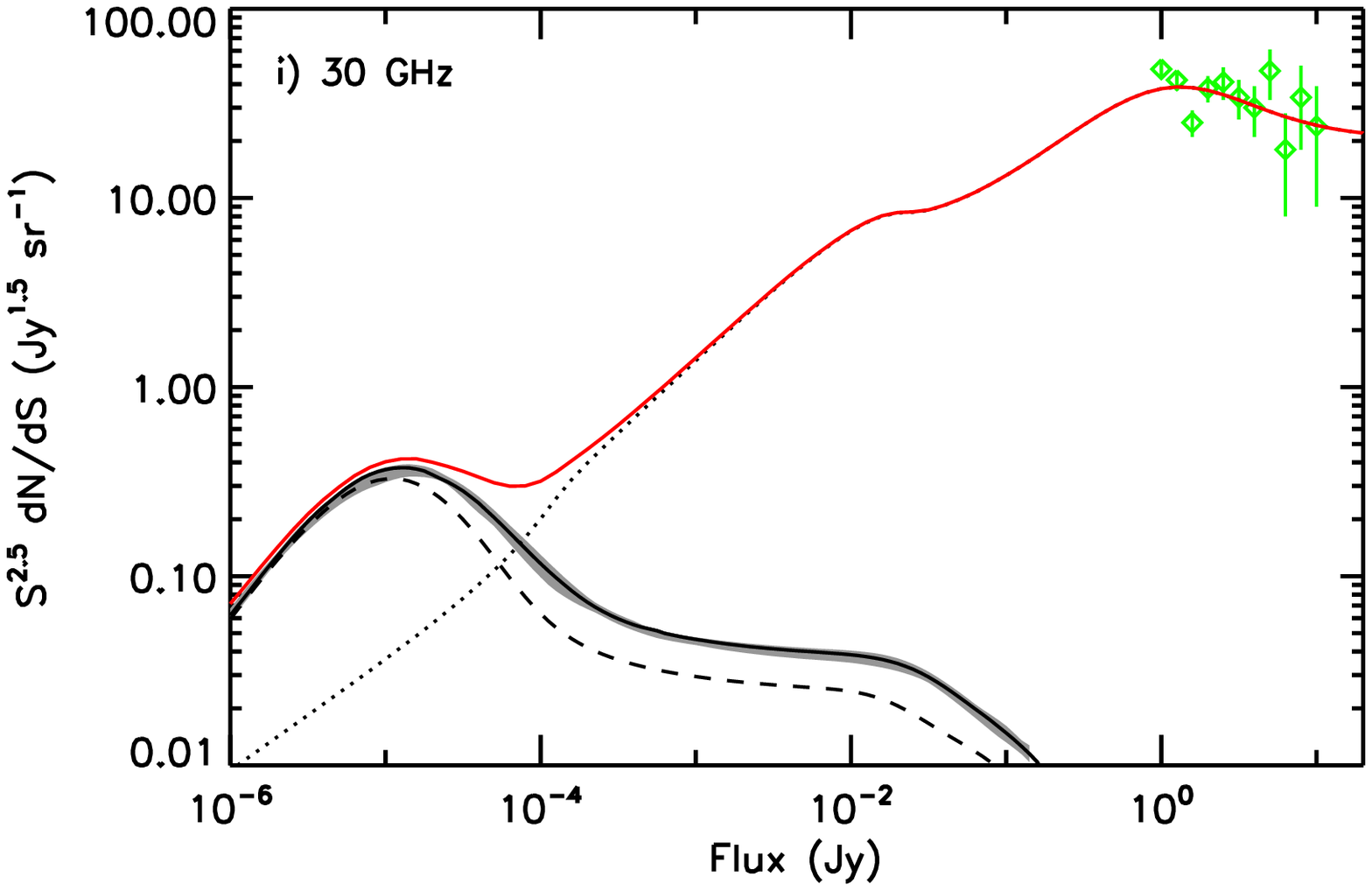} \\
\includegraphics[width=0.32\textwidth]{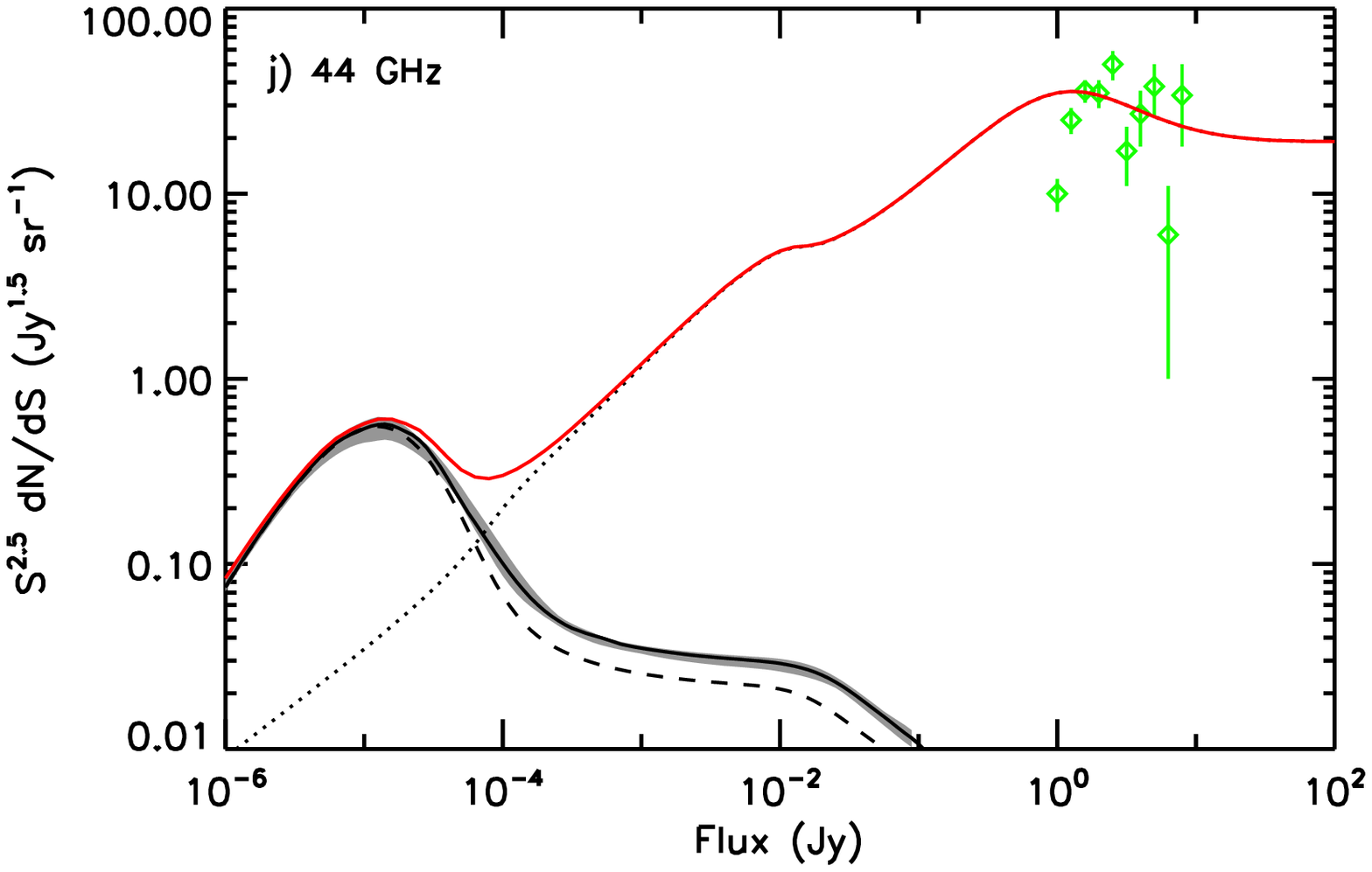} & \includegraphics[width=0.32\textwidth]{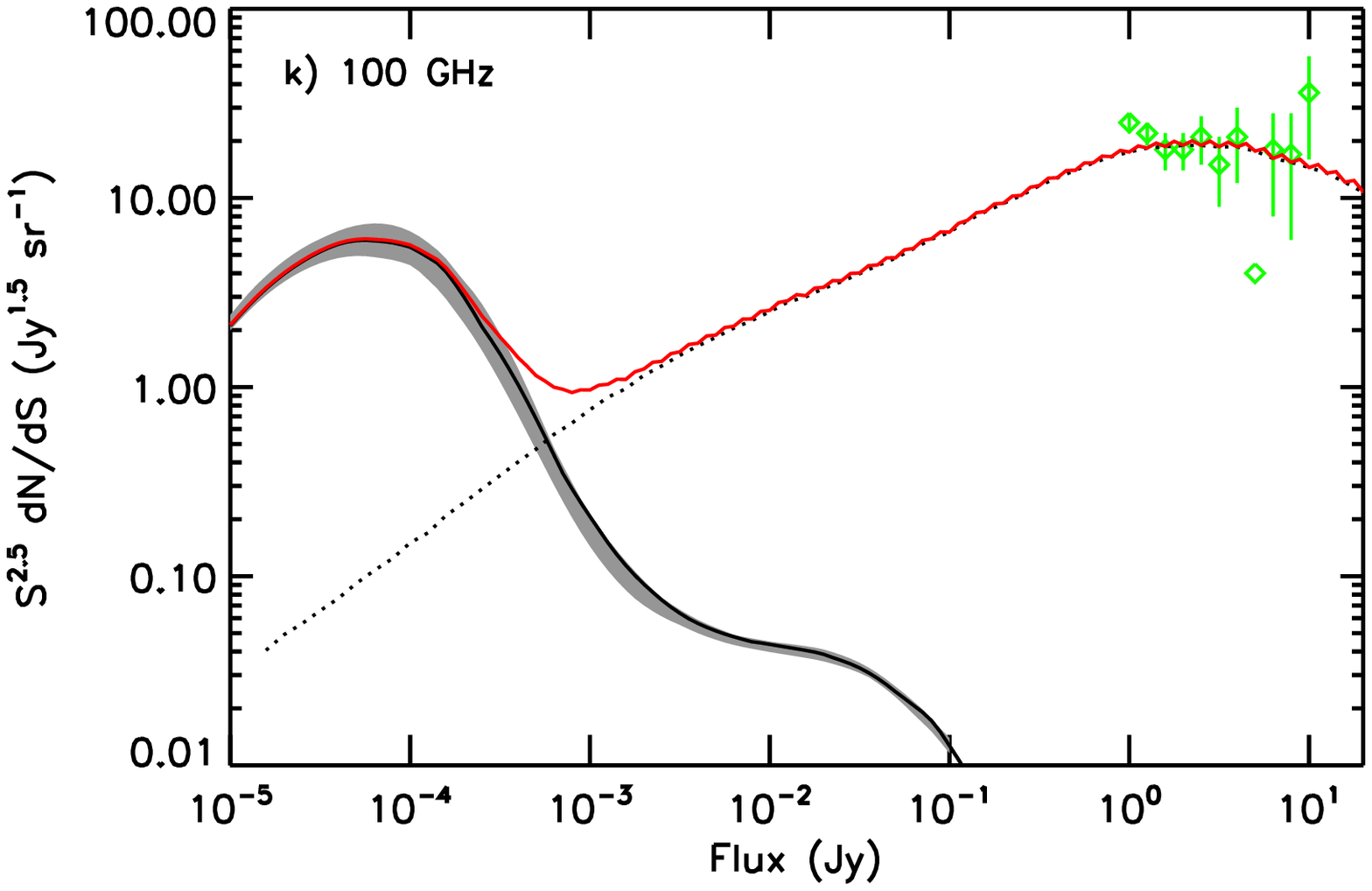} & \includegraphics[width=0.32\textwidth]{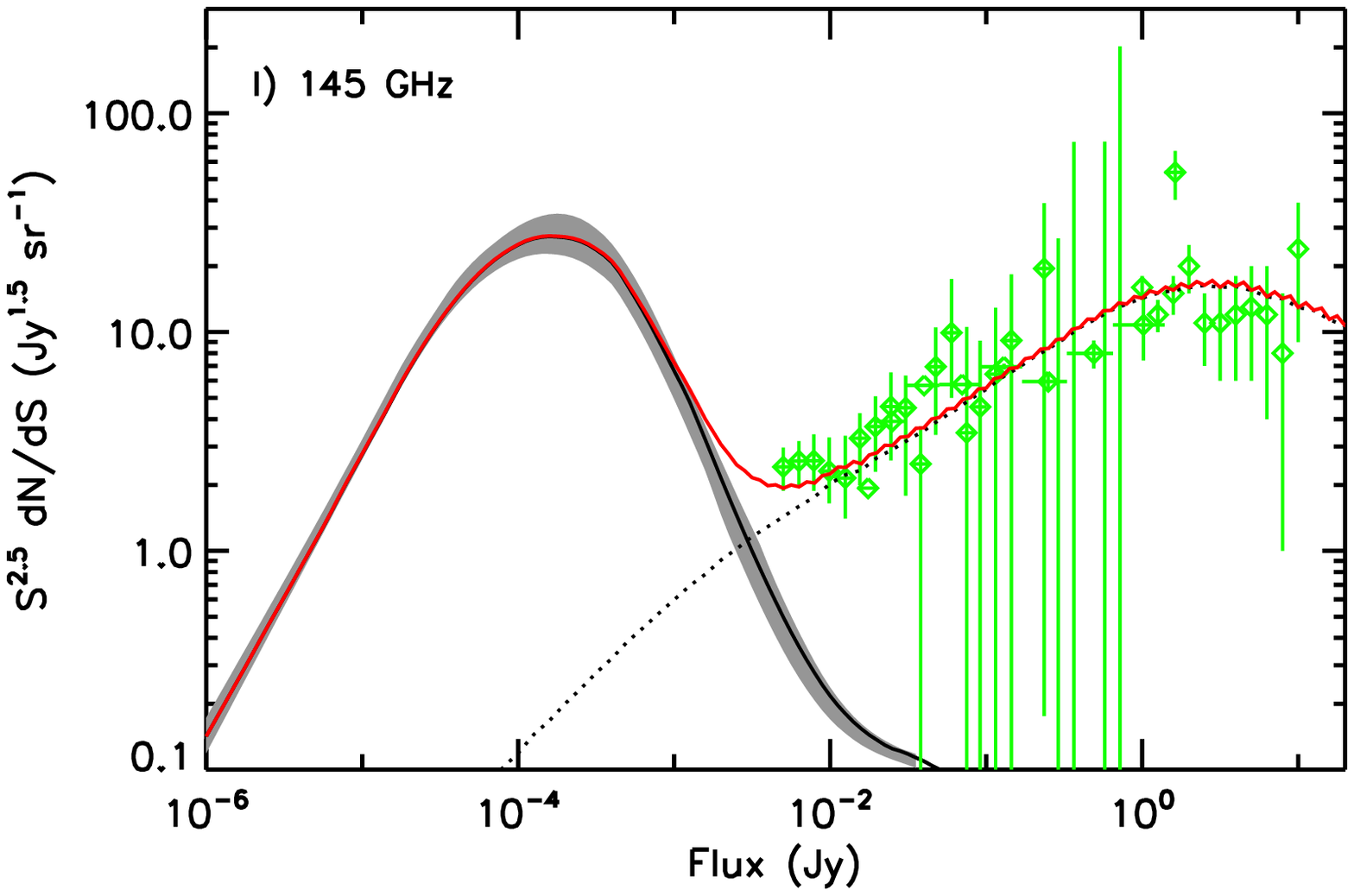} \\
\includegraphics[width=0.32\textwidth]{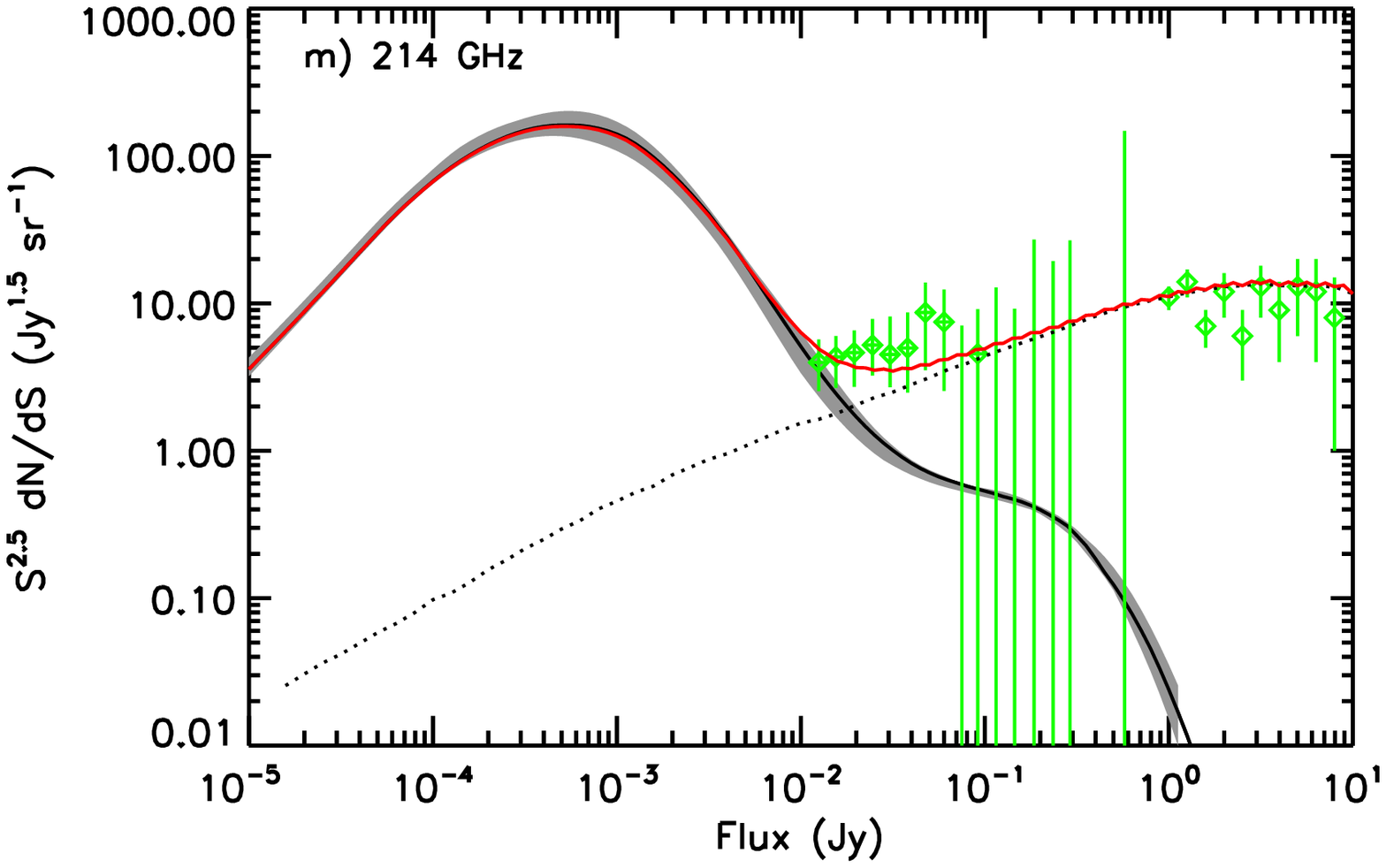} & \includegraphics[width=0.32\textwidth]{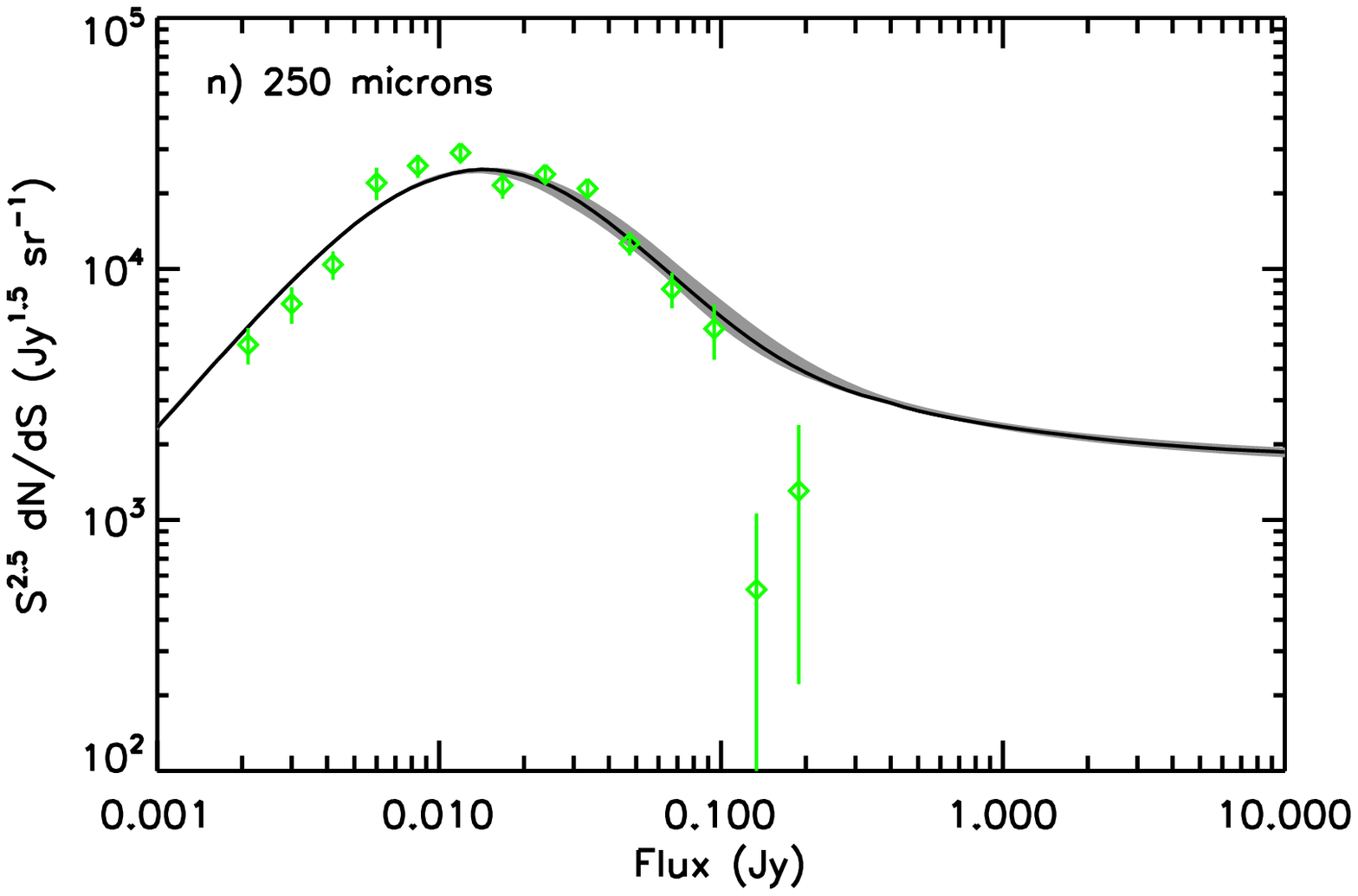} & \includegraphics[width=0.32\textwidth]{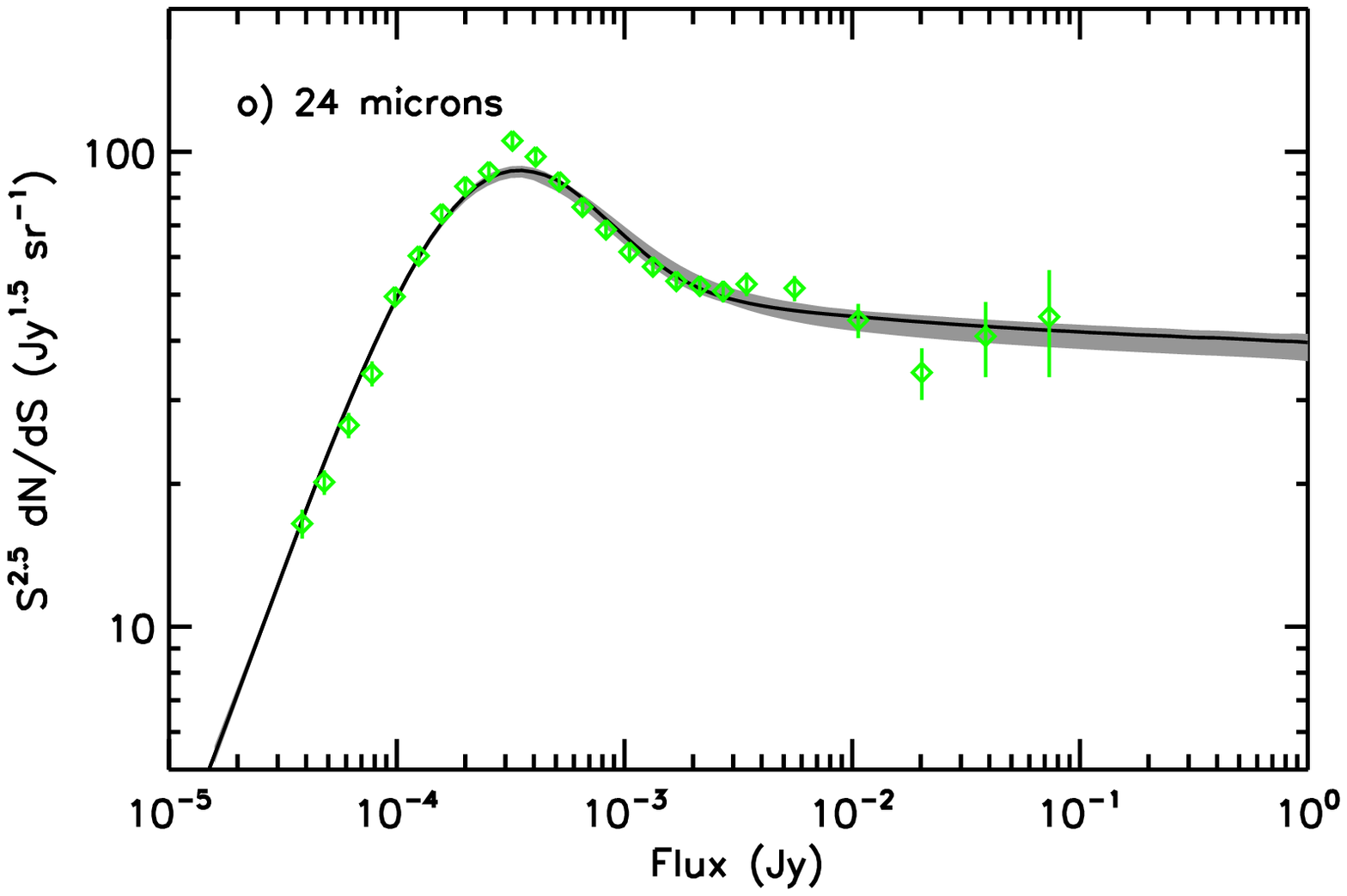}
\end{tabular}}
\caption{Differential extragalactic number counts (green triangles) for $\nu = 408$~MHz (a), 610~MHz (b), 1.4~GHz (c), 2.7~GHz (d), 4.8-5~GHz (e), 8.4~GHz (f), 15~GHz (g), 20~GHz (h), 30~GHz (i), 44~GHz (j), 100~GHz (k), 143-145-148~GHz (l), 214-217~GHz (m), 250~$\mu$m (n), and 24~$\mu$m (o). The counts are a compilation from several surveys (see Tab. \ref{table_references} for references). The black solid and dashed lines are the predictions of our evolution model with and without spinning dust emission, respectively, in the case of a constant far-IR/radio flux ratio (see case $A$ of Sect. \ref{section_synchrotron}) and using the LF-parameters of the {\it mean model} of \citet{Bethermin2011}. The grey areas show the 1-$\sigma$ error calculated from the MCMC fit of the LF by \citet{Bethermin2011}. The dotted lines are the predictions of evolution models for radio-loud AGN as described in Sect. \ref{section_AGN} \citep{deZotti2005, Massardi2010, Tucci2011}. The red lines show the sum of the contribution of radio-loud AGN and star-forming galaxies when spinning dust emission is included to the SED templates.}
\label{counts} 
\end{figure*}

\begin{table}
\label{table_references}
\centering
\caption{References for the extragalactic number counts presented in Figs. \ref{counts}, \ref{counts_qz} and \ref{counts_add_high_z}.}
\begin{tabular}{ll}
\hline
\hline
Frequency & References \\
\hline
408 MHz    & \citet{Gervasi2008} and references therein \\
610 MHz    & \citet{Gervasi2008} and references therein \\
1.4 GHz    & \citet{Gervasi2008} and references therein \\
2.7 GHz    & \citet{Gervasi2008} and references therein \\
4.8 GHz    & \citet{Vernstrom2011} \\
5 GHz      & \citet{Gervasi2008} and references therein \\
8.4 GHz    & \citet{Tucci2011} \\
15.7 GHz   & \citet{AMI2011} \\
20 GHz     & \citet{Tucci2011} \\
100 GHz    & \citet{PlanckDole2011} \\
143 GHz    & \citet{PlanckDole2011} \\
145 GHz    & \citet{SPT2010} \\
148 GHz    & \citet{Marriage2011} \\
214 GHz    & \citet{SPT2010} \\
217 GHz    & \citet{PlanckDole2011} \\
250 $\mu$m & \citet{Bethermin2012} \\
24 $\mu$m  & \citet{Bethermin2010} \\
\hline
\end{tabular}
\end{table}

\begin{figure}[!t]
\centerline{
\includegraphics[width=0.4\textwidth]{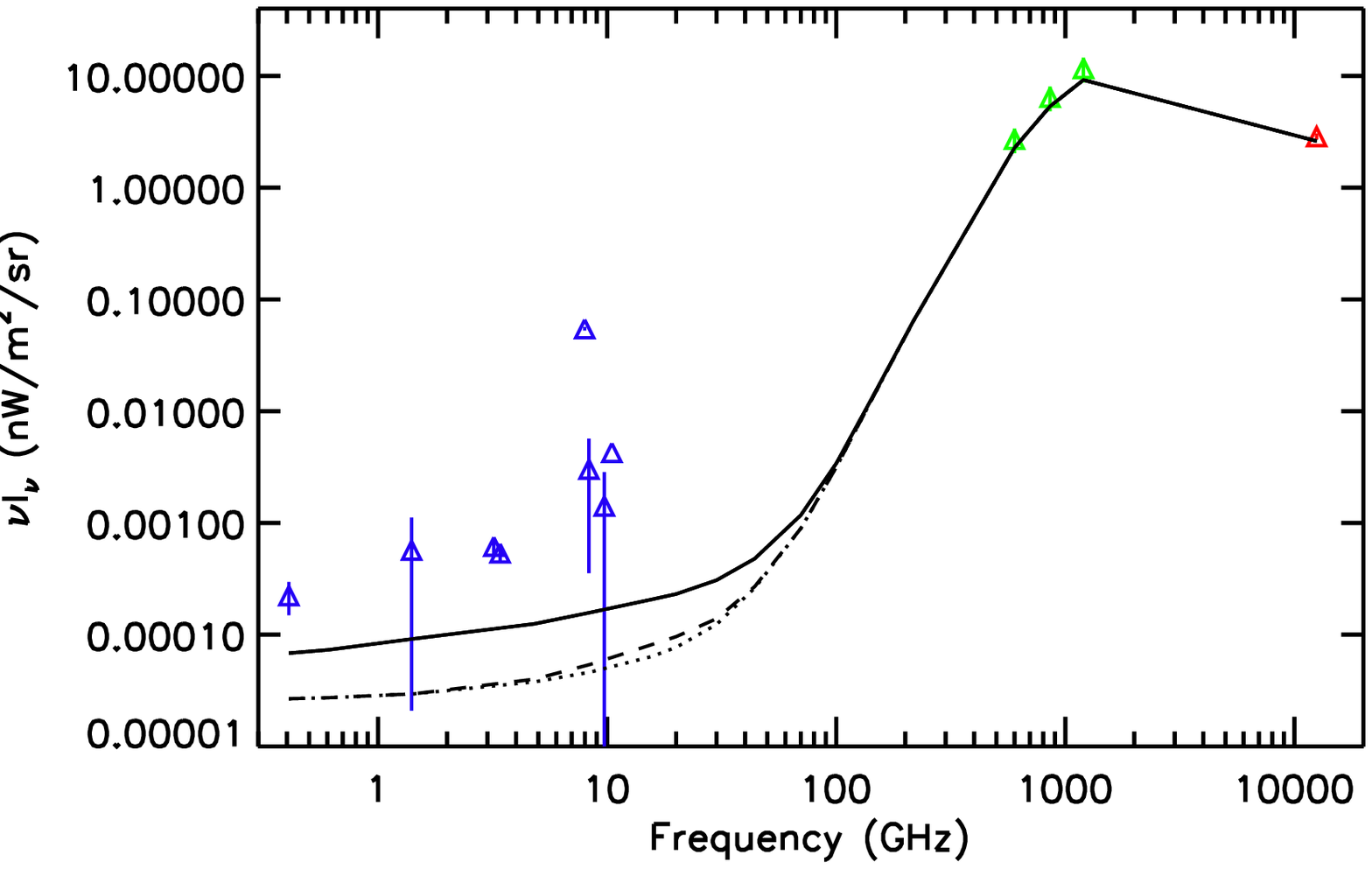}}
\caption{Cosmic background spectrum predicted by our model for dusty star-forming and normal spiral galaxies without (dotted line) and with (dashed line) spinning dust emission. The solid line shows the spectrum when the contribution of radio-loud AGN is included \citep{deZotti2005, Massardi2010, Tucci2011}. The blue triangles show direct measurements of the CRB obtained by ARCADE2 \citep{Fixsen2011} and older surveys reprocessed using the model developed by the ARCADE2 team to measure the CRB (408 MHz and 1.4~GHz). The green triangles show direct measurements of the CIB at 250, 350, and 500~$\mu$m using FIRAS data processed as in \citet{Lagache1999}. The red triangle shows the CIB measured in the MIPS 24~$\mu$m band by \citet{Bethermin2010}.}
\label{CIB} 
\end{figure}

\begin{table}
\label{CIB_table}
\centering
\caption{Surface brightness of the cosmic background for case $A$ (first column), contribution of AGN (second column), and contribution of the IR galaxies (third column).}
\begin{tabular}{cccc}
\hline
\hline
Frequency &  Total                & AGN                   & IR galaxies           \\
(GHz)     & (nW/m$^2$/sr)         & (nW/m$^2$/sr)         & (nW/m$^2$/sr)         \\
\hline
0.408     & 6.83$\times$10$^{-5}$ & 4.16$\times$10$^{-5}$ & 2.67$\times$10$^{-5}$ \\
0.61      & 7.32$\times$10$^{-5}$ & 4.60$\times$10$^{-5}$ & 2.73$\times$10$^{-5}$ \\
1.4       & 9.19$\times$10$^{-5}$ & 6.24$\times$10$^{-5}$ & 2.96$\times$10$^{-5}$ \\
5         & 1.25$\times$10$^{-4}$ & 8.51$\times$10$^{-5}$ & 3.99$\times$10$^{-5}$ \\
8.4       & 1.58$\times$10$^{-4}$ & 1.04$\times$10$^{-4}$ & 5.42$\times$10$^{-5}$ \\
15.7      & 2.07$\times$10$^{-4}$ & 1.26$\times$10$^{-4}$ & 8.16$\times$10$^{-5}$ \\
20        & 2.31$\times$10$^{-4}$ & 1.35$\times$10$^{-4}$ & 9.62$\times$10$^{-5}$ \\
30        & 3.07$\times$10$^{-4}$ & 1.68$\times$10$^{-4}$ & 1.39$\times$10$^{-4}$ \\
44        & 4.79$\times$10$^{-4}$ & 2.09$\times$10$^{-4}$ & 2.69$\times$10$^{-4}$ \\
70        & 1.18$\times$10$^{-3}$ & 2.81$\times$10$^{-4}$ & 8.94$\times$10$^{-4}$ \\
100       & 3.42$\times$10$^{-3}$ & 2.57$\times$10$^{-4}$ & 3.17$\times$10$^{-3}$ \\
143       & 1.31$\times$10$^{-2}$ & 2.98$\times$10$^{-4}$ & 1.28$\times$10$^{-2}$ \\
217       & 6.44$\times$10$^{-2}$ & 3.55$\times$10$^{-4}$ & 6.40$\times$10$^{-2}$ \\
\hline
\end{tabular}
\end{table}

The predictions for the differential extragalactic number counts in case $A$ (constant $q_{70}$, Sect. \ref{section_synchrotron}) are presented in Fig. \ref{counts}. The contributions of both dusty star-forming galaxies and AGN are included \citep{deZotti2005, Massardi2010, Tucci2011}. For the dusty star-forming galaxies, Fig. \ref{counts} also presents the uncertainties on the counts calculated from the MCMC fit of the LF by \citet{Bethermin2011} in the case of the SEDs described in Sects. \ref{section_free_free}, \ref{section_synchrotron}, and \ref{section_spinning}.

In agreement with \citet{Massardi2010}, we find that the low-flux bump of the counts in the radio (610~MHz and 1.4~GHz) can be accounted for by standard star-forming galaxies. Moreover, our model can explain {\it all} the galaxy number counts from the mid-IR (24~$\mu$m) to the radio (408~MHz). Our model shows that star-forming galaxies account for 84.5\% of the extragalactic number counts at 1.4~GHz for a flux of 50~$\mu$Jy and 77.5\% for 100~$\mu$Jy.   

The CRB predicted by our model can be calculated by summing the galaxy number counts. The results are shown in Fig.~\ref{CIB} and the corresponding surface brightness from 408~MHz to 217~GHz are listed in Tab. \ref{CIB_table}. The contribution of star-forming galaxies to the CRB reaches 39.1\% at 408~MHz, 37.2\% at 610~MHz, and 32.2\% at 1.4~GHz. These values are much higher than those found by \citet{Ponente2011}, who used a model based on an empirical relation between the 1.4~GHz luminosity and the star formation rate density (SFRD). This discrepancy could be caused by the fact that these authors used a SFRD/radio conversion ratio that is the same for all redshifts. This ratio should rather be weighted by the luminosity of the galaxies that dominate the averaged SFRD at a given redshift. According to \citet{Lagache2004}, the IR output energy is dominated by galaxies with bolometric luminosities around $L_{IR} \sim 8\times 10^{10} \; L_{\odot}$ at $z = 0$, whereas it is dominated by galaxies with $L_{IR} \sim 3\times 10^{12} \; L_{\odot}$ at $z = 2$. Furthermore, \citet{Ponente2011} do not consider spinning dust emission and assume slopes of $\beta = -2.7$ for the radio emission, respectively. For local galaxies, the measurements made taking into account the presence of spinning dust emission show that it is steeper with $\beta \sim -3$.

As can be seen from Fig.~\ref{counts}, the contribution of spinning dust to the galaxy number counts is significant at 8.44, 15.7, 20, and 30~GHz. However, deeper surveys would be required to validate the model in this frequency range as the star-forming galaxies dominate the counts only for $S \sim 10 \; \mu$Jy, where no data are available. Then, spinning dust emission accounts for about 10 to 20\% of the cosmic background produced by star-forming galaxies in the microwave (14.8\% at 8.44~GHz, 21.6\% at 15.7~GHz, 19.8\% at 20~GHz, and 11.1\% at 30~GHz). However, this represents only less than 10\% of the total background when AGN are included (5.1\% at 8.44~GHz, 8.5\% at 15.7~GHz, 8.3\% at 20~GHz, and 5\% at 30~GHz).

Finally, Fig.~\ref{CIB} shows that even if our model is able to reproduce all the galaxy number counts and the cosmic background in the mid- and far-IR, it cannot explain the excess radio emission reported by ARCADE2 \citep{Fixsen2011}.

\subsection{Evolution of $q_{70}$ with redshift (case $B$)}
\label{section_qz}

\begin{figure}[!t]
\centerline{
\begin{tabular}{c}
\includegraphics[width=0.4\textwidth]{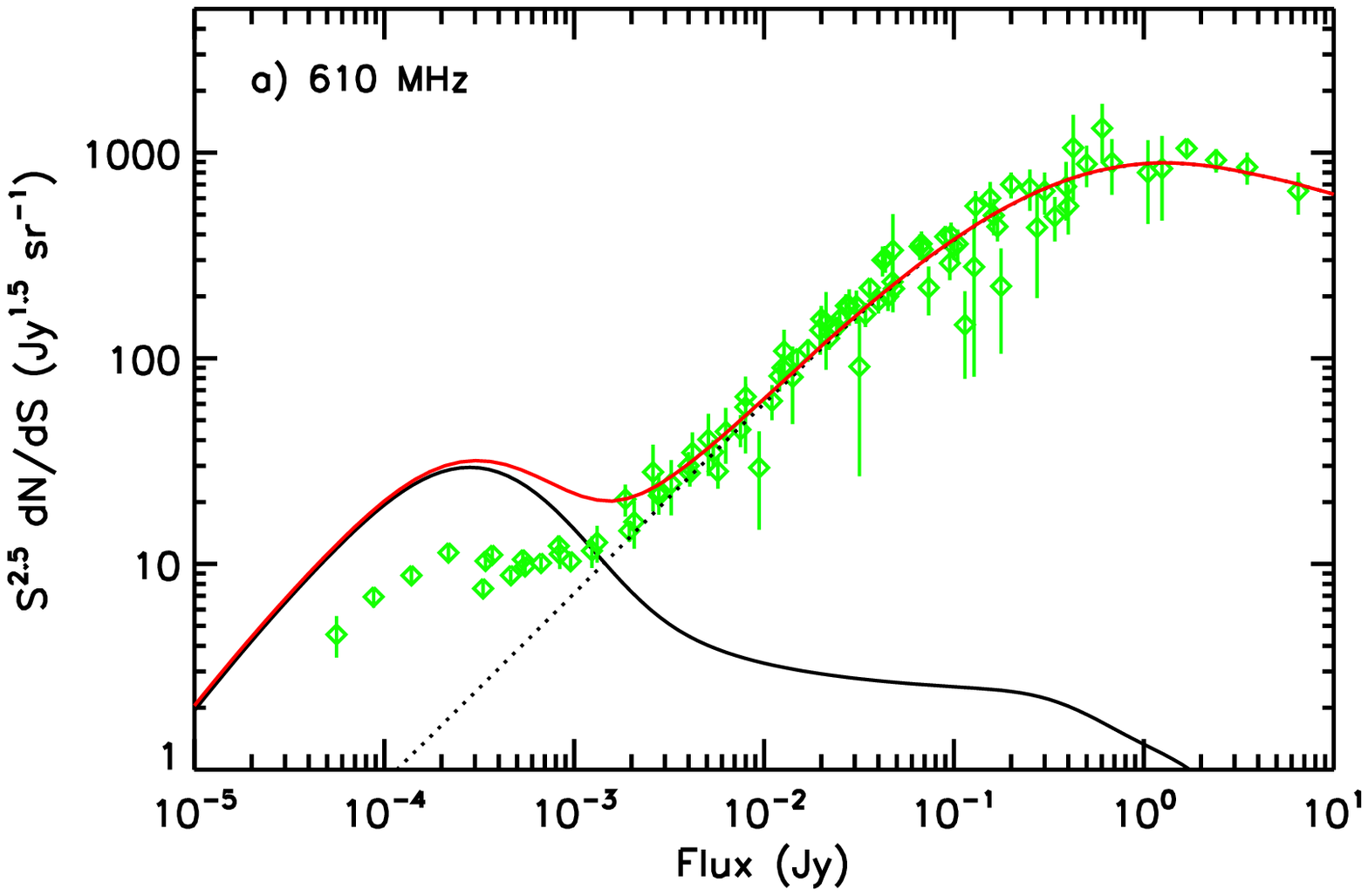} \\
\includegraphics[width=0.4\textwidth]{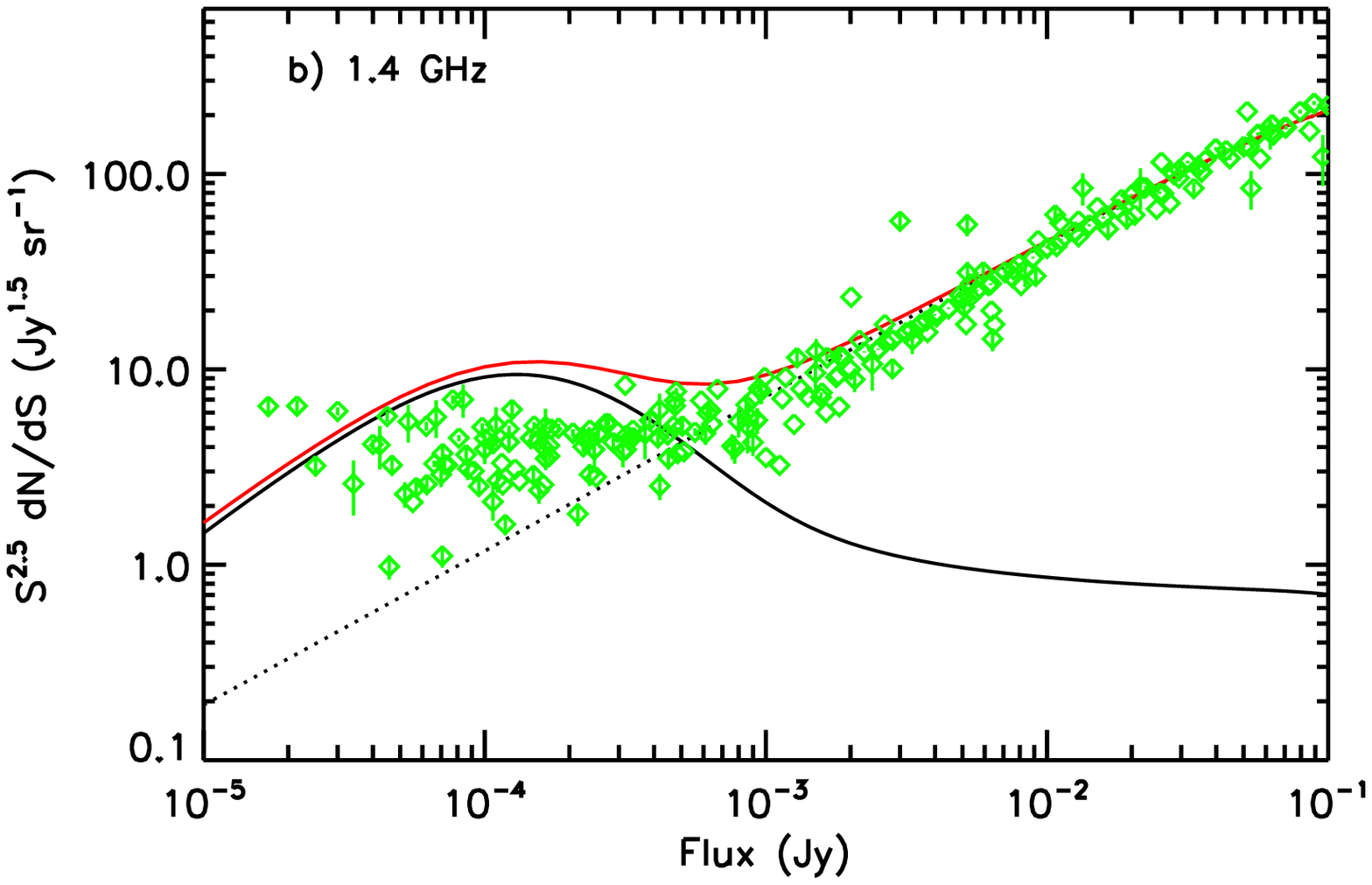}
\end{tabular}}
\caption{Differential extragalactic number counts (green triangles) for $\nu = 610$ MHz (a) and 1.4 GHz (b), in the case of a variable far-IR/radio flux ratio $q_{70}(z) = 2.15 - 0.72\log(1+z)$ (see case $B$ in Sect. \ref{section_synchrotron}).}
\label{counts_qz} 
\end{figure}

The predictions of the differential extragalactic number counts for case $B$ ($q_{70}$ decreasing with redshift, see Sect. \ref{section_synchrotron}) are presented in Fig. \ref{counts_qz} at 610 MHz and 1.4 GHz.  As expected, allowing $q_{70}$ to vary increases the radio emissivity of star-forming galaxies and leads to an increase in the CRB of about 70\%, which is however still not enough to explain ARCADE2 data: ARCADE measurements are still greater than the model predictions by a factor of 5 at 408~MHz and of 11.4 at 1.4~GHz. Then, case $B$ yields an overestimate of the galaxy counts in the radio range (Fig. \ref{counts_qz}). This suggests that the far-IR/radio flux ratio should rather be constant or at least decrease less rapidly with redshift as suggested by \citet{Ivison2010}. According to the compilation of data made by \citet{Ponente2011}, see their Fig. 1, it seems reasonable to consider that the far-IR/radio ratio is constant at first order for $z \lesssim 3.5$ \citep{Bell2003, Murphy2009, Ivison2010b, Ivison2010, Michalowski2010, Sargent2010, Bourne2011}. 

We test the case of star-forming galaxies having the properties of case $A$ for $z \leqslant 4$ and an enhanced radio emission above. The value of $q_{70}(z \geqslant 4)$ needs to be equal to $\sim 0$ for the model predictions to reach the ARCADE2 radio excess. However, this again yields to an overestimate of the galaxy counts in the radio range, even worse than for case $B$. As a result, even if the far-IR/radio flux ratio decreases only for high redshift, it cannot explain ARCADE2 excess without being in contradiction with extragalactic number counts.

\subsection{An additional population of faint galaxies at high redshift?}
\label{section_add_high_z}

\begin{figure}[!t]
\centerline{
\begin{tabular}{c}
\includegraphics[width=0.4\textwidth]{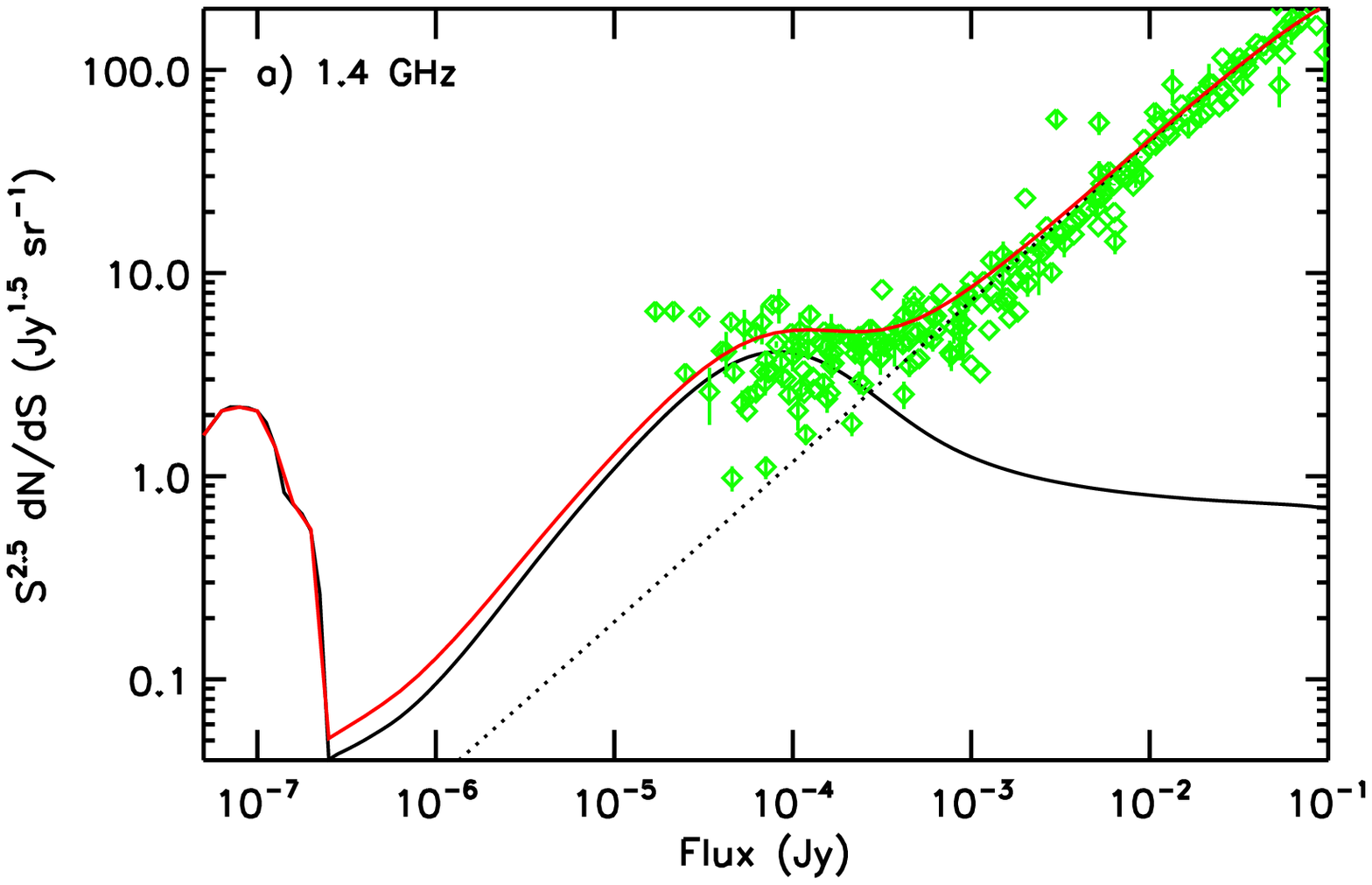} \\
\includegraphics[width=0.4\textwidth]{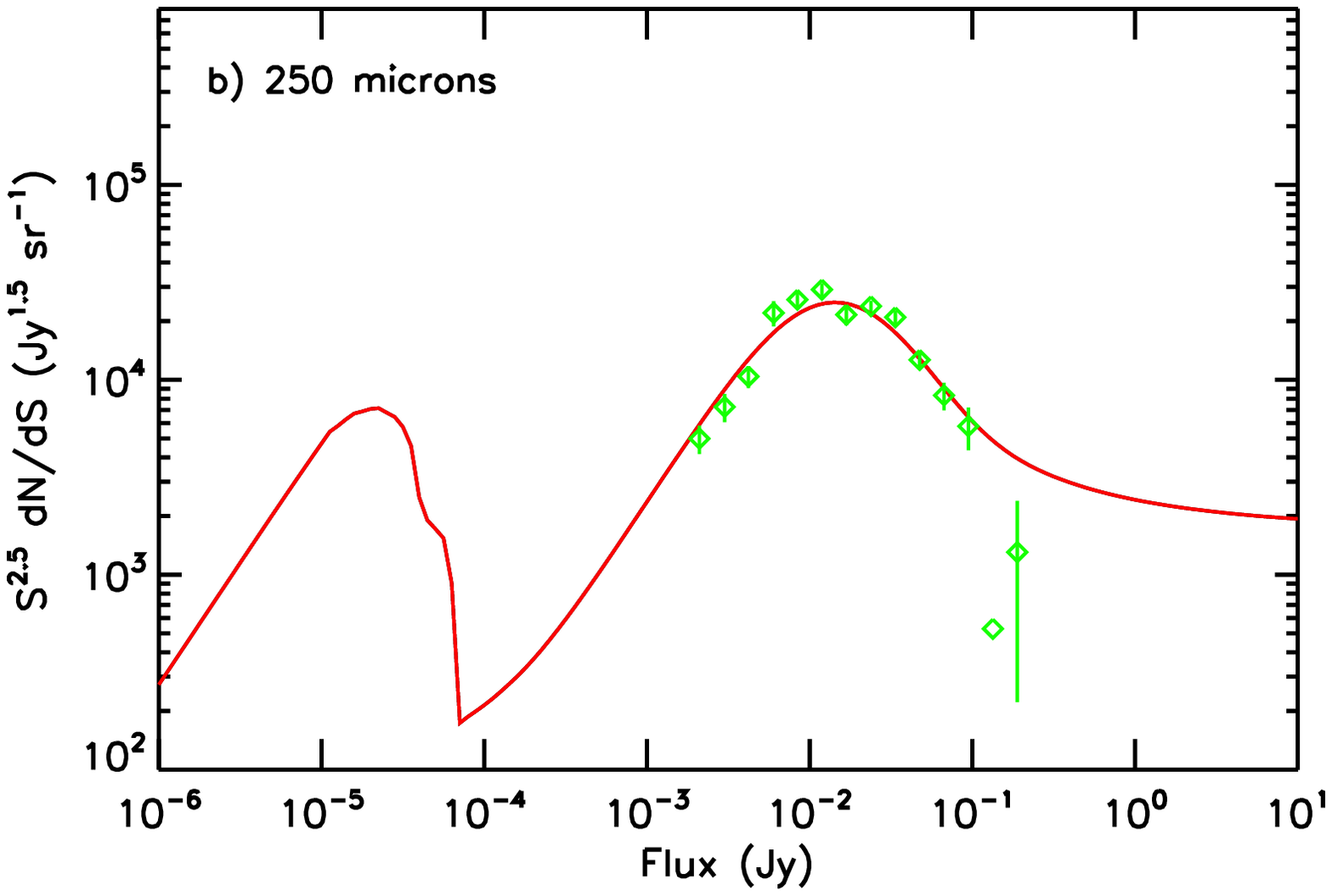} \\
\includegraphics[width=0.4\textwidth]{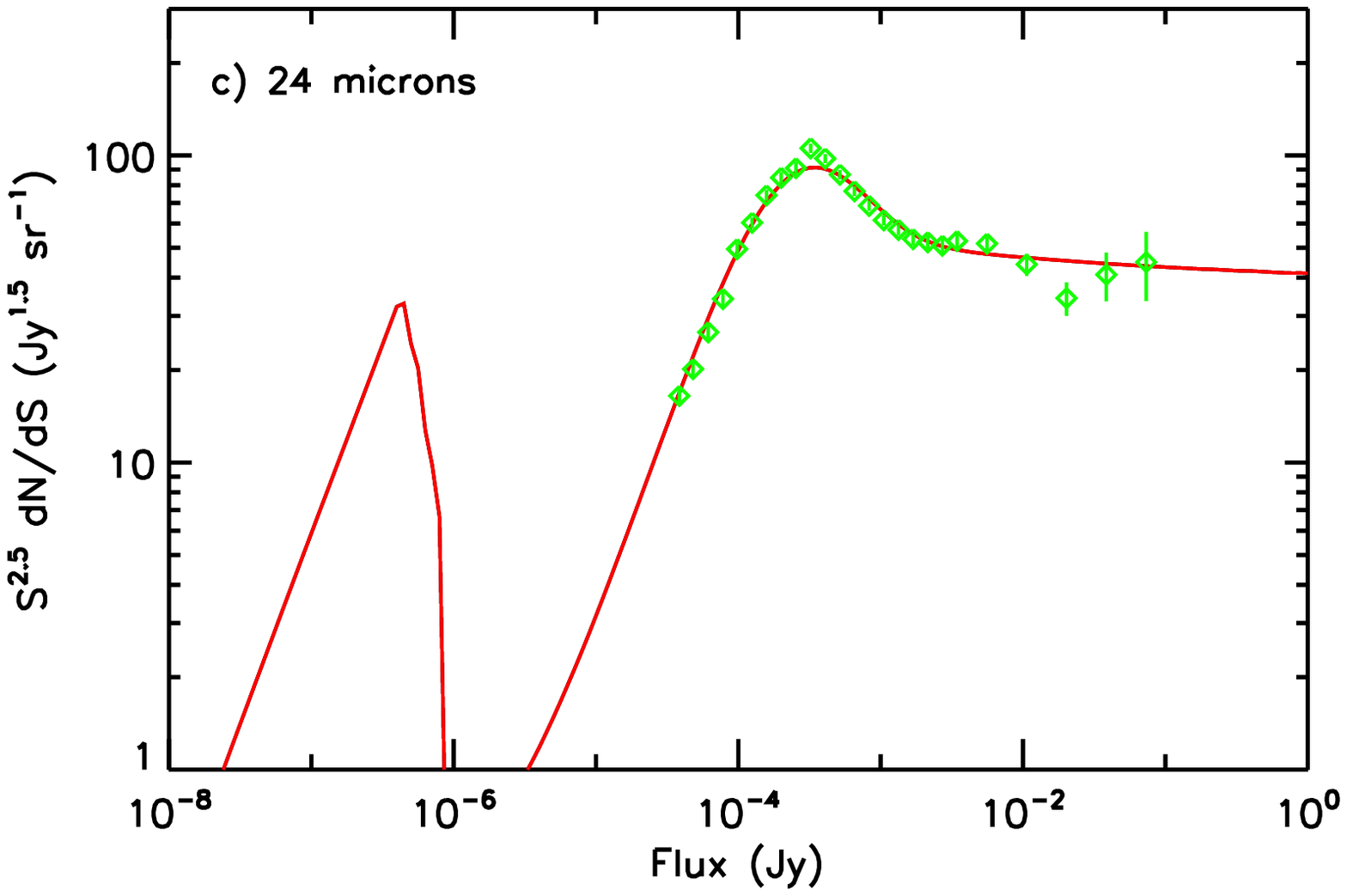} \\
\includegraphics[width=0.4\textwidth]{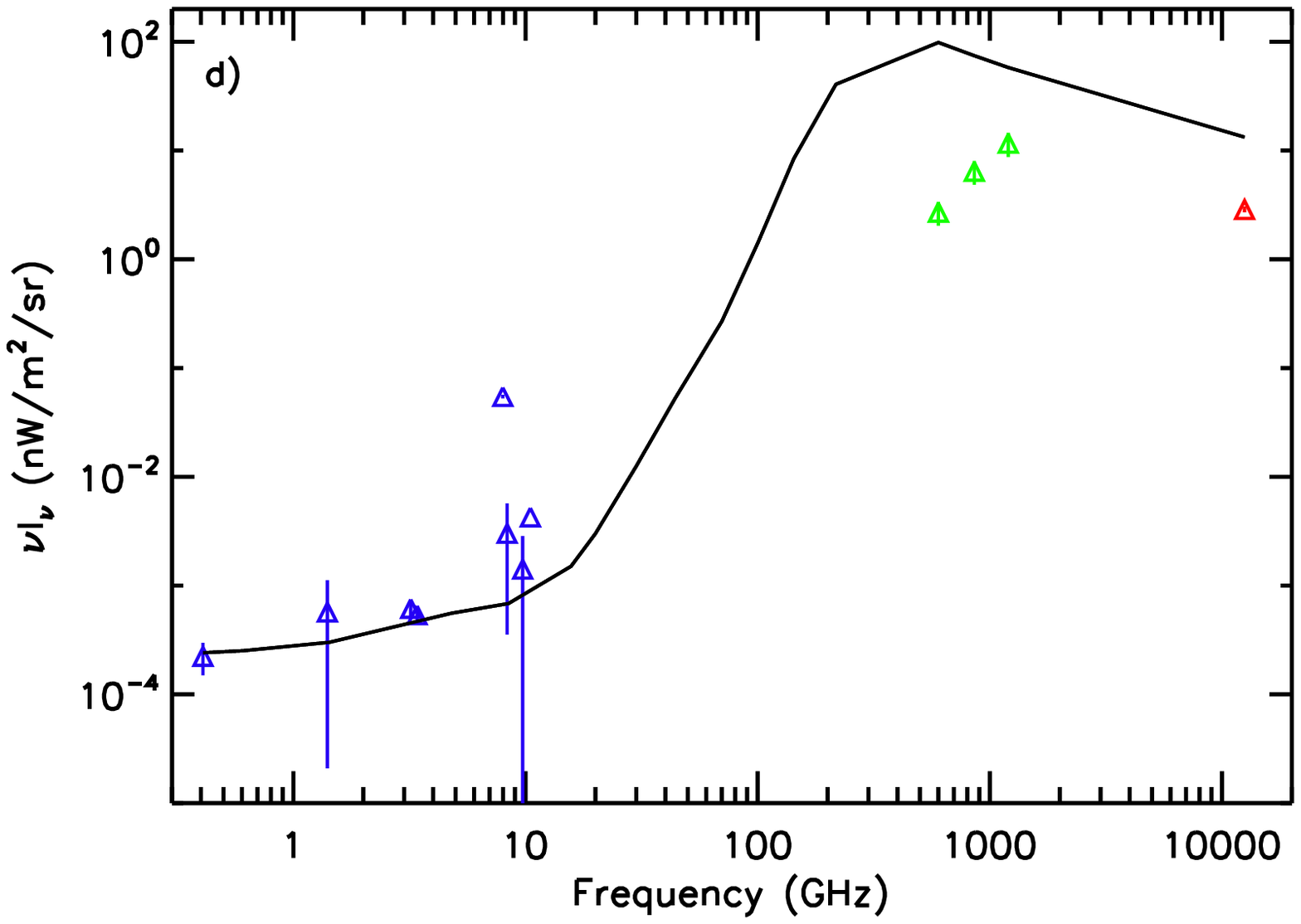} 
\end{tabular}}
\caption{(a), (b), (c) Differential extragalactic number counts at 1.4 GHz, 250 $\mu$m, and 24 $\mu$m (linestyles and symbols are the same as in Fig. \ref{counts}) predicted by our model when a population of faint galaxies is added at high redshift (see Sect. \ref{section_add_high_z} for details). (d) Corresponding cosmic background. Linestyles are the same as in Fig. \ref{CIB}.}
\label{counts_add_high_z} 
\end{figure}

If we suppose that the radio excess observed by ARCADE2 originates in star-forming galaxies, then the only remaining way to increase the CRB is to consider an additional population of faint IR galaxies at high redshift. It is important to consider faint and not bright galaxies not to disrupt the galaxy number counts and also because we do not expect a large amount of bright galaxies to be undetected even at high redshift.

We use the galaxy templates of case $A$ (Sect. \ref{section_synchrotron}) and start with the luminosity functions of \citet{Bethermin2011}. We multiply them by a constant for $4 \leqslant z \leqslant 6$ and for $L_{IR} \leqslant L_{thres}$. The lowest $L_{thres}$ allowing to reproduce the ARCADE2 radio excess is $L_{thres} = 10^{11} \; L_{\odot}$. For lowest threshold luminosities, it is not possible to fit simultaneously the background surface brightness at 408 MHz and 5 GHz. Higher threshold luminosities predict too many galaxy number counts at low flux ($\leqslant 1$~mJy) that we ruled out by current observational constraints. The model predictions for $L_{thres} = 10^{11} \; L_{\odot}$ are presented in Fig. \ref{counts_add_high_z}. The increase in the luminosity function required to match the ARCADE2 excess overproduces the CIB by a factor of 4.6 at 24~$\mu$m, 8.5 at 250~$\mu$m, 11.6 at 350~$\mu$m, and 21.5 at 500~$\mu$m. We conclude that a population of faint galaxies at high redshift able to account for the radio excess reported by ARCADE2 is ruled out by direct measurements of the CIB.

\section{Conclusions}
\label{section_conclusions}

The aim of this paper was to check if star-forming galaxies could be responsible for the radio excess reported by \citet{Fixsen2011} in the CRB with ARCADE2 data. To do so, using up-to-date detailed star-forming and radio galaxy evolution models, we considered three scenarios: $(i)$ contribution of spinning dust emission, $(ii)$ variation with redshift of the far-IR/radio flux ratio, and $(iii)$ additional population of faint IR galaxies at high redshift.

$(i)$ Our model, including the contribution of synchrotron, free-free, and spinning dust emissions, and considering a constant far-IR/radio flux ratio, accounts for all the extragalactic number counts from the mid-IR to the radio (24 $\mu$m to 408 MHz). It also accounts for the cosmic background in the mid- and far-IR. However, it does not reproduce the ARCADE2 excess. We note that our model agrees with the findings of \citet{Zannoni2008}, \citet{Gervasi2008b}, and \citet{Tartari2008}, who could explain the TRIS observations with no need for an excess in the radio range (600 MHz, 820 MHz, and 2.5 GHz). Our model also accounts for the CRB intensity as estimated by \citet{Wall1970} at 320 and 707 MHz. Finally, we confirm that radio-loud AGN dominate the counts for $S_{\nu} \geqslant 1$ mJy \citep{Ibar2008}. 

$(ii)$ In order to increase the radio emission of the star-forming galaxies, we considered a possible decrease in the far-IR/radio flux ratio, $q_{70}$, with redshift. The steepest relation available in the litterature does not yield an acceptable fit to the extragalactic number counts at 610 MHz and 1.4 GHz. We conclude that varying $q_{70}$ with redshift cannot explain the ARCADE2 radio excess. These two results suggest that the far-IR/radio flux ratio should rather be constant, or at least decrease less rapidly with redshift than what was measured by \citet{Seymour2009}.

$(iii)$ Finally, we tested the hypothesis of an additional population of faint star-forming galaxies at high redshift. We show that the amount of faint galaxies required to explain the ARCADE2 radio excess leads to a predicted CIB much higher than what is actually observed at 250, 350, and 500 $\mu$m.

Consequently, we exclude the star-forming galaxies as the origin of the ARCADE2 radio excess. Furthermore, \citet{Singal2010} investigated the contribution to the CRB of radio-quiet quasars, radio supernovae, emission from hot gas in galaxy clusters, and diffuse emission from regions far from galaxies. They found that they cannot account for more than about 10\% of the excess. Considering our results, and the results of \citet{Singal2010}, we conclude that if the radio excess reported by \citet{Fixsen2011} is astrophysical, it has to have a Galactic origin or to originate in a new kind of radio sources (with no mid- to far-IR counterparts) or emission mechanism still to be discovered. 
 
\acknowledgements{We thank our anonymous referee for useful comments. We also thank M. B\'ethermin for his help and making his model available. We thank H. Dole, M. Zannoni, M. Tucci, and N. Seymour for useful discussions and making their data and models available. We are also grateful to D. Dicken for his help concerning the English language! N.Y. acknowledges the support of a CNES post-doctoral research grant.}

\bibliography{biblio}

\end{document}